\documentclass[preprintnumbers,amsmath,amssymb,aps,prx,longbibliography,floatfix,lengthcheck,superscriptaddress]{revtex4-2}
\usepackage{graphicx}
\usepackage{braket}
\usepackage{upgreek}
\usepackage[colorlinks=true,citecolor=blue]{hyperref}
\usepackage{xcolor}
\usepackage{orcidlink}

\newcommand{\unit}[2]{\mbox{\ensuremath{#1}\,#2}}

\begin{document}

\title{Motional ground-state cooling of single atoms in state-dependent optical tweezers}

\author{C. H\"{o}lzl\orcidlink{0000-0002-2176-1031}}
\author{A. G\"{o}tzelmann\orcidlink{0000-0001-5527-5878}}
\author{M. Wirth\orcidlink{0009-0007-6940-7916}}
\affiliation{5. Physikalisches Institut and Center for Integrated Quantum Science and Technology, Universit\"{a}t Stuttgart, Pfaffenwaldring 57, 70569 Stuttgart, Germany}
\author{M. S. Safronova\orcidlink{0000-0002-1305-4011}}
\affiliation{Department of Physics and Astronomy, University of Delaware, Newark, Delaware 19716, USA}
\affiliation{Joint Quantum Institute, National Institute of Standards and Technology and the University of Maryland, College Park, Maryland 20742, USA}
\author{S. Weber\orcidlink{0000-0001-9763-9131}}
\affiliation{Institute for Theoretical Physics III and Center for Integrated Quantum Science and Technology, Universit\"{a}t Stuttgart, Pfaffenwaldring 57, 70569 Stuttgart, Germany}
\author{F. Meinert\orcidlink{0000-0002-9106-3001}}
\affiliation{5. Physikalisches Institut and Center for Integrated Quantum Science and Technology, Universit\"{a}t Stuttgart, Pfaffenwaldring 57, 70569 Stuttgart, Germany}
\date{\today}

%%%%%%%%%%%%%%%%%%%%%%%%%  Abstract  %%%%%%%%%%%%%%%%%%%%%%%%%
\begin{abstract}
Laser cooling of single atoms in optical tweezers is a prerequisite for neutral atom quantum computing and simulation. Resolved sideband cooling comprises a well-established method for efficient motional ground-state preparation, but typically requires careful cancellation of light shifts in so-called magic traps. Here, we study a novel laser cooling scheme which overcomes such constraints, and applies when the ground-state of a narrow cooling transition is trapped stronger than the excited state. We demonstrate our scheme, which exploits sequential addressing of red sideband transitions via frequency chirping of the cooling light, at the example of $^{88}$Sr atoms and report ground-state populations compatible with recent experiments in magic tweezers. The scheme also induces light-assisted collisions, which are key to the assembly of large atom arrays. Our work enriches the toolbox for tweezer-based quantum technology, also enabling applications for tweezer-trapped molecules and ions that are incompatible with resolved sideband cooling conditions.
\end{abstract}

\maketitle

\section{Introduction}
Quantum control of individual atoms trapped in optical tweezers has seen a very rapid development in the last years \cite{Kaufman2021}, which has opened routes for applications such as quantum computing and simulation \cite{Browaeys2020,Bluvstein2022,Graham2022}, precision metrology \cite{Madjarov2019,Young2020}, or ultracold chemistry \cite{Cheuk2020,Cairncross2021}. The quest for high-fidelity operations on the internal states of the atoms, for example, logic gate operations or optical clock interrogation, also requires cooling of the external motion, ideally down to the quantum mechanical ground-state of the tweezer trap \cite{Thompson2013,Kaufman2012}. Efficient ground-state preparation is also key for assembling Hubbard-type lattice models atom by atom with optical tweezers \cite{Young2022,Spar2022}, or for realizing ultrafast quantum gate protocols via resonant F\"orster interactions \cite{Chew2022}.

Large motional ground-state occupation is typically achieved using well-established sideband-resolved cooling protocols \cite{Thompson2013,Kaufman2012}, which more recently also became available for alkaline-earth(-like) atom arrays exploiting their narrow intercombination transitions \cite{Cooper2018,Norcia2018,Saskin2019}. Sideband laser cooling, however, poses tight constraints on the trapping condition, as it requires careful cancellation of differential light shifts for the internal atomic states involved in the cooling cycle, a situation referred to as magic trapping. Consequently, many of the applications mentioned above, including optical tweezer-based atomic clocks \cite{Madjarov2019,Young2020} or novel concepts for qubit implementations in gate-based quantum processors \cite{PatentFineStructureQubit,Pagano2022}, are incompatible with such constraints. Novel laser cooling mechanisms that work at more general conditions thus not only expand the toolbox for neutral atom quantum technology, but may also find applications in controlling optically trapped molecules or even ions \cite{Anderegg2019,Caldwell2020,Schneider2010}.

\begin{figure}[!ht]
\centering
	\includegraphics[width=\columnwidth]{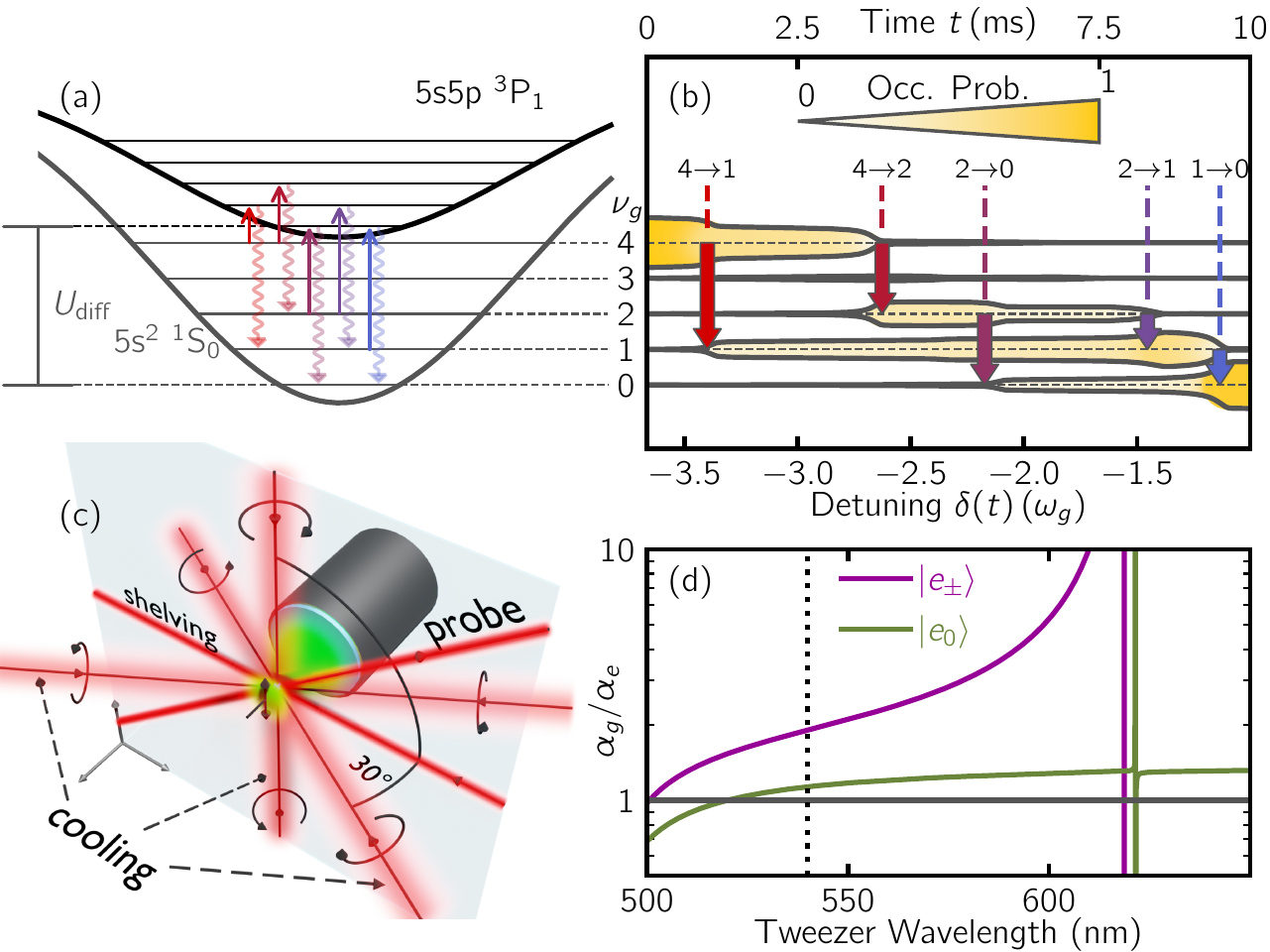}
	  \caption{Chirp-cooling in state-dependent optical tweezers.
    (a) Radial tweezer potential for the $|g\rangle=|^1S_0\rangle$ and
    $|e\rangle=|^3P_1\rangle$ states of the cooling transition. The
    state-dependent trap depth enables transfer of excited motional states into the
    trap ground-state via strong red sideband transitions, reducing $\nu_g$ by
    $\Delta \nu>1$ vibrational quanta (arrows denote excitation and subsequent
    spontaneous decay). The potential curves are offset vertically so that
    $U_{\rm{diff}}$ indicates the (positive) light shift of the electronic
    transition between the motional ground-states from the free-space resonance. (b)
    Numerical simulations of the cooling dynamics, demonstrating ground-state
    transfer when the laser detuning $\delta$ (measured from the trap shifted
    resonance at $U_{\rm{diff}}$) is ramped from a large red detuning towards
    $\delta \approx -\omega_g$. The thickness of the yellow areas is proportional to
    the population in $|\nu_g\rangle$. Arrows indicate population transfer when the
resonance condition for the sideband transitions drawn in (a) is matched. (c) Experimental realization: single $^{88}$Sr atoms are trapped in an optical tweezer (green), using a high-NA objective. The polarization of the tweezer light is indicated by the arrow in the NA-cone. Chirp-cooling is realized with three pairs of counter-propagating, circularly polarized $689 \, \rm{nm}$ MOT beams (cooling). Additional laser beams (probe and shelving) are used for sideband thermometry. (d) Ratio of the AC-polarizability (and hence of the trap depths) between $|^1S_0\rangle$ and the light-shifted substates $|e_0\rangle$ and $|e_\pm\rangle$ of $|^3P_1\rangle$ as a function of trapping wavelength. The dotted line indicates the wavelength used in the experiment.}
	\label{Fig1}
\end{figure}

In this paper, we demonstrate a method for motional ground-state cooling at the
example of single trapped $^{88}$Sr atoms which is applicable in the generic
situation of sizable differential light shifts between the relevant atomic
states. The strategy relies on a frequency chirp of the cooling light to quench
the population of initially occupied motional states towards the trap ground
state. A detailed theoretical proposal and analysis thereof has been reported recently in Ref. \cite{Berto2021}. Optimizing the cooling parameters, we measure ground-state populations
exceeding \unit{80}{\%} along the radial (more strongly confined) tweezer axis. At the same time, the cooling protocol efficiently removes pairs of atoms from the trap via light-assisted collisions resulting in a final $50\%$ filling probability with exactly one atom, prerequisite for the systematic assembly of large atom arrays \cite{Endres2016,Barredo2016}.

\section{Ground-state cooling scheme for state-dependent tweezers}
We consider an atom with two internal electronic states $|g\rangle$ and $|e\rangle$ trapped in the Gaussian-shaped potential formed by an optical tweezer (Fig.~\ref{Fig1}(a)). Sufficiently close to the trap bottom, the atom is harmonically confined with state-dependent oscillator frequencies $\omega_g$ and $\omega_e$, respectively. Standard sideband cooling (Raman- or single-photon scheme) requires equal AC-polarizabilities $\alpha_e$ and $\alpha_g$ for the electronic levels, yielding identical ladders of motional states independent of the atom's internal state. Provided a narrow laser transition between $|g\rangle$ and $|e\rangle$ with linewidth $\gamma \ll \omega_g$ (\textit{festina lente} regime, see Refs. \cite{Santos2000,Urvoy2019} for its importance in all-optical cooling schemes to Bose-Einstein condensation), one can then cool the atomic motion in the trap by setting the laser to a fixed frequency resonant with the first red sideband, i.e. detuned by $-\omega_g$ from the free-space transition frequency. Magic trapping ( $\alpha_e = \alpha_g$) ensures, that the condition for addressing the red sideband is independent of the harmonic oscillator level.

Such persistent cooling conditions are no longer given when the trapping
potential is state-dependent ($\alpha_e \neq \alpha_g$). Only very recently, it
has been demonstrated that efficient cooling into the trap ground-state can
still be realized with a fixed laser frequency for $\alpha_g/\alpha_e < 1$, i.e.
when the excited state $|e\rangle$ is trapped stronger than the ground-state
$|g\rangle$ \cite{Covey2019,Urech2022}. Here, we investigate the opposite case,
$\alpha_g/\alpha_e > 1$, for which cooling with a fixed frequency cannot work.
This can be seen in Fig.~\ref{Fig1}(a), which depicts the situation we study
in our experiment. Specifically, we employ the narrow ($\gamma = 2 \pi \times
7.4 \, \rm{kHz}$) $^1S_0$ to $^3P_1$ intercombination transition at a laser
wavelength of about $689 \, \rm{nm}$ for in-trap cooling. It is convenient to
define the laser detuning $\delta$ with respect to the trap-shifted resonance
condition $U_{\rm{diff}}/\hbar$ (see Fig.~\ref{Fig1}(a)) for driving the
electronic transition with the atom in the motional ground-state, i.e.
$|g,\nu_g=0\rangle$ to $|e,\nu_e=0\rangle$. Here, $\nu_g (\nu_e)$ is the
harmonic oscillator quantum number of the ground (excited) state vibrational
ladder. In our scenario where $\omega_e < \omega_g$, the resonance condition to
drive the red sideband $\delta \approx (\nu_g -1)\omega_e- \nu_g \omega_g$ now depends on $\nu_g$.
An attempt to cool the lowest-lying vibrational excitations requires to set
$\delta$ close to the red sideband condition for $\nu_g=1$, i.e. $\delta \approx -\omega_g$.
Such a laser frequency, however, causes heating of higher-lying vibrational
states. Consequently, cooling with a fixed frequency would not succeed. This
problem can be resolved by using a time-dependent chirp of $\delta$ which
compensates for the $\nu_g$-dependence of the condition to address red sidebands
one after the other. Such a frequency chirp then dissipates motional quanta without concurrent heating,
since the protocol assures that higher vibrational states are first transferred to lower energy, before states closer to the trap bottom are addressed (see also Ref. \cite{Berto2021} for a recent proposal). Note that a fixed laser frequency provides an effective repulsive energy cap in the trap, which was exploited in Ref.~\cite{Cooper2018} to prevent atom loss during imaging and which was interpreted by a classical Sisyphus effect.

Before we turn to the experimental results, we briefly analyze the chirp-cooling
approach numerically. To this end, we compute the quantum dynamics of a
harmonically confined and laser-coupled (Rabi frequency $\Omega$) two-level atom
in 1D including state-dependent trapping. The time evolution is obtained by
integrating the Liouville-von Neumann equation for the density matrix with a
finite basis set of oscillator levels accounting for decay of the excited state
(decay rate $\gamma$) via the Lindblad operators (see Appendix \ref{SM:chirp_num}).
Results for typical
experimental parameters are shown in Fig.~\ref{Fig1}(b) for a linear (\unit{10}{ms} long) ramp of
$\delta$ from $\delta_i / \omega_g = - 3.7$ to $\delta_f /
\omega_g = - 1$ and for the atom initially prepared in
$|\nu_g=4\rangle$. The data reveals that chirping allows for efficient transfer
into the motional ground-state. For the chosen parameters $\left[ (\omega_g,\omega_e,\Omega)=2\pi \unit{(150,110,20)}{kHz}\right]$, the final
ground-state population is $>94 \%$. Cooling occurs due to resonant
addressing of various red sidebands during the chirp, coupling states of
different motional quantum numbers $\nu_g > \nu_e$ (arrows in Fig.~\ref{Fig1}(a) and (b)). Note that in contrast to magic trapping conditions, where couplings between different oscillator levels with $\Delta \nu = |\nu_g - \nu_e| > 1$ are strongly suppressed for tight confinement by the Lamb-Dicke effect, here, they play a vital role in the cooling dynamics due to direct wavefunction overlaps $\langle \nu_g | \nu_e \rangle$ between levels of equal parity \cite{Taieb1994}. The simulations also reveal optimal ground-state population when $\delta_f \approx - \omega_g $, i.e. when the laser frequency chirp ends near the resonance condition for driving the first red sideband from $|\nu_g=1\rangle$.
Note that the simulation parameters for Fig. \ref{Fig1}(b) are chosen in a way that the underlying mechanism of the cooling scheme is well visible and are not optimized for highest ground-state transfer.

\section{Cooling and thermometry}
Our experiments start with loading a single optical tweezer with wavelength
$\lambda = 539.91\, \rm{nm}$ and a waist of
$\unit{564(5)}{nm}$ from a $^{88}$Sr magneto-optical trap (MOT) operated on the
$^1S_0$ to $^3P_1$ intercombination line (Fig.~\ref{Fig1}(c))
(see Appendix \ref{SM:tweezer_loading}).
After loading, the tweezer is typically occupied by more than one atom in the
electronic ground-state $^1S_0$ (trap depth $\approx 0.5 \, \rm{mK}$). At the trapping wavelength, the tweezer
potential for $^1S_0$ is deeper than for the excited $^3P_1$ state of the
cooling transition, i.e. realizing the situation $\alpha_g/\alpha_e > 1$
discussed above. More precisely, we perform all experiments at nominally zero
magnetic field and with a linearly polarized tweezer. In that case, the three
magnetic substates of the $^3P_1$ level ($m_J=0,\pm1$) are shifted by
the AC-Stark interaction with the trap light. We label the AC-Stark-shifted
eigenstates $|e_0\rangle$, $|e_-\rangle$, and $|e_+\rangle$. The latter two are
energetically degenerate (see Ref. \cite{Cooper2018} and Appendix \ref{SM:diffac}). The wavelength dependence of the
ratio $\alpha_g/\alpha_e$ for all three levels is shown in Fig.~\ref{Fig1}(d).
This allows us to perform experiments using two different transitions with vastly
different values $\alpha_g/\alpha_e$ ($\sim 1.13$ for $|e_0\rangle$ and $\sim
1.90$ for $|e_\pm\rangle$).

We start investigating the cooling dynamics on the transition to $|e_\pm
\rangle$, which exhibits the stronger differential light shift, measured to be
$U_{\rm{diff}}^{|e_\pm\rangle} / \hbar = 2 \pi \times \unit{5.50(5)}{MHz}$ for
the tweezer's optical power of about $\unit{1.70(2)}{mW}$ set throughout
this work. The cooling protocol starts by switching on the $689 \, \rm{nm}$ MOT
beams with an initial detuning $\delta_i = - 2 \pi \times 4.1 \, \rm{MHz}$. We estimate the Rabi frequency to about $2 \pi \times 50 \, \rm{kHz}$. The beams are kept on for $100 \, \rm{ms}$, during which the laser frequency is ramped linearly to a variable final detuning $\delta_f$.

\begin{figure}[!ht]
\centering
	\includegraphics[width=\columnwidth]{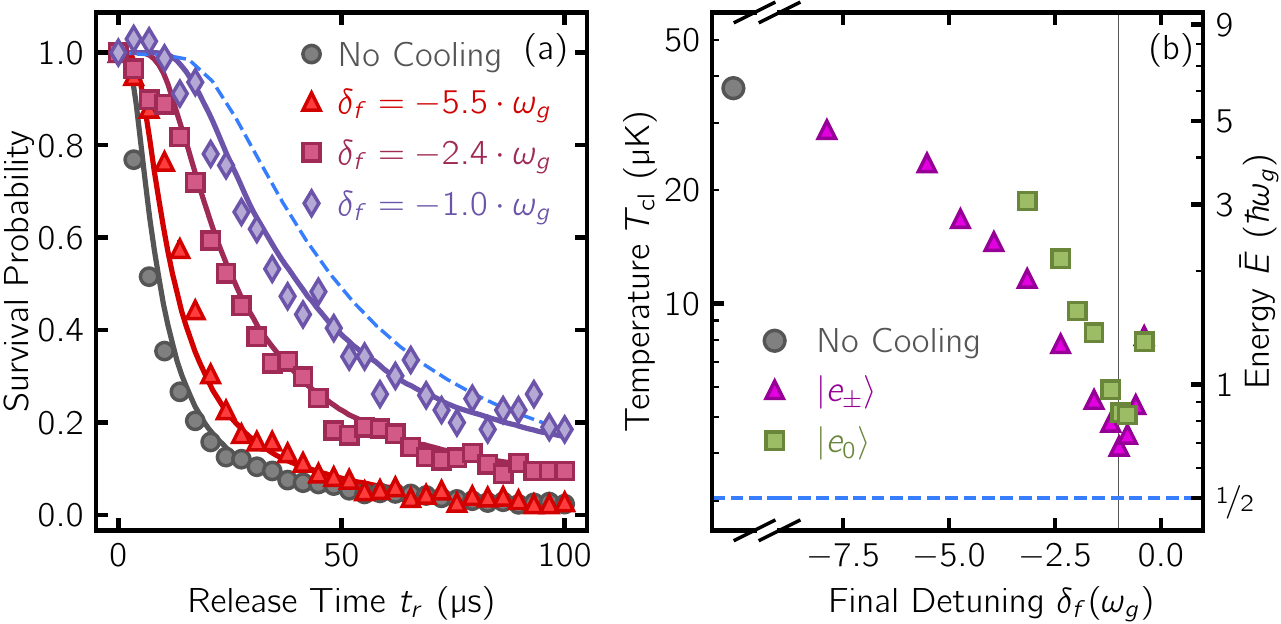}
  \caption{Thermometry after chirp-cooling via release-and-recapture.
    (a) Atom-survival probability as a function of the release time $t_r$
    is shown for three different values of $\delta_f$ as indicated
    (triangles, squares, diamonds) after cooling on the transition to $|e_\pm
    \rangle$, and compared to a measurement with no cooling applied (circles).
    Solid lines are fits of a classical particle trajectory simulation to the
    data to extract the temperature. The blue-dashed line shows the prediction of a
    quantum mechanical simulation for an atom in the motional ground-state
    (see Appendix \ref{SM:rr_num}).
    (b) Temperature $T_{\rm{cl}}$ extracted from classical trajectory simulations of data as shown in
    (a) as a function of the detuning $\delta_f$ at the end of the frequency chirp.
    Triangles (squares) are results for cooling on the transition to $|e_\pm
    \rangle$ ($|e_0 \rangle$). The circle denotes the temperature without cooling
    applied. The vertical line indicates the resonance condition ($\delta_f = -\omega_g$)
    for the lowest red sideband $|g,\nu_g=1\rangle \rightarrow |e,\nu_e=0\rangle$.
    At this point the minimal temperature is achieved. The dashed line depicts the
    temperature-equivalent $T_{\rm{gs}}$ of the radial zero-point motion energy in the trap.
    Error bars in all figures show one standard deviation, and are mostly smaller than the data points.
  }
	\label{Fig2}
\end{figure}

The temperature after the ramp is measured via the release-and-recapture
technique \cite{Tuchendler2008}. Briefly, the trap is turned off diabatically
for a variable release time $t_r$ before it is suddenly switched on again. We
then image the atoms on the ${^1}S_0$-${^1}P_1$ transition at 461nm by collecting
fluorescence photons on a sCMOS camera. From the photon signal on the camera, we deduce the survival probability
of a single atom in the trap (see Appendix \ref{SM:imaging} and below for more details).  Exemplary data of such measurements is shown in Fig.~\ref{Fig2}(a) for different values of $\delta_f$ and compared to data taken without the cooling protocol (circles). A slower decay of the measured survival probability with $t_r$ is indicative for lower temperature, as hotter atoms escape faster from the trap volume. The results reveal a reduction of temperature with decreasing $|\delta_f|$, i.e. when the frequency is chirped closer to the light-shifted resonance at $U_{\rm{diff}}^{|e_\pm\rangle} / \hbar$.

To extract a classical temperature $T_{\rm{cl}}$ at the end of the chirp from datasets as shown in
Fig.~\ref{Fig2}(a), we fit classical Monte-Carlo simulations of the
release-and-recapture sequence to the data, assuming a thermal energy
distribution with mean energy $\bar{E} = k_B T_{\rm{cl}}$ in each spatial direction (solid lines) \cite{Tuchendler2008}. Results of this analysis are
plotted in Fig.~\ref{Fig2}(b) as a function of $\delta_f$ (triangles). The data
reveal a minimum temperature of $\unit{4.17(6)}{$\upmu$K}$ in the vicinity of
$\delta_f \approx -\omega_g$. Throughout this work $\omega_g = \unit{2 \pi \times 126(5)}{kHz}$, as measured via sideband spectroscopy (see below). Compared to the data taken without cooling, we achieve a reduction in temperature by about one order of magnitude. Chirping further down in $|\delta_f|$ again leads to heating, as the cooling light approaches the resonance condition for driving the carrier transition from the ground-state ($\delta=0$).

The minimal measured temperature is found close to the temperature-equivalent of
the zero-point motion energy of the radial tweezer ground-state $T_{\rm{gs}} =
\hbar \omega_g / 2 k_B = 3.04 \, \rm{\upmu K}$ (dashed line), indicating
sizeable radial ground-state population. Since the classical analysis does not
account for the zero-point motion in the trap, we also analyze the experimental data
with the lowest-measured temperature quantum mechanically. As the
release-and-recapture method is only weakly dependent on the axial tweezer
direction, it is sufficient to model the recapture probability along the radial
direction. To this end, we time-evolve the wavefunctions of the first few states
of a 2D harmonic oscillator numerically. After time $t_r$ of free expansion, the
Gaussian tweezer potential is added to deduce the probability of recapture
(see Appendix \ref{SM:rr_num}). The result of this analysis for the 2D ground-state (dashed line in
Fig.~\ref{Fig2}(a)) is already close to the lowest-temperature experimental data
($\delta_f = -\omega_g$). Next, the analysis is extended to a thermal
distribution of the first few 2D harmonic oscillator states. A fit of this model
to the data for $\delta_f = -\omega_g$ yields $82(3) \%$ ground-state population
along one radial direction. To demonstrate the robustness of the chirp-cooling,
we finally repeat the measurements on the transition to $|e_0 \rangle$, for
which the differential light shift $U_{\rm{diff}}^{|e_0\rangle} / \hbar = 2 \pi
\times 1.30(2) \, \rm{MHz}$ is much weaker. We obtain very similar results (squares in Fig.~\ref{Fig2}(b)), and attribute the slightly higher minimal temperature to the higher sensitivity required for tuning $\delta_f$.

\begin{figure}[!t]
\centering
	\includegraphics[width=\columnwidth]{fig3.pdf}
	\caption{(a) Scheme for sideband thermometry via shelving into metastable
    states. The atom in the ground-state $^1S_0$ is excited to the $^3P_1$
    ($|e_0\rangle$) state by a short probe laser pulse. The population
    transferred to $^3P_1$ is hidden from the imaging cycle (${^1}S_0$-${^1}P_1$)
    by optically pumping into the metastable $^3P_0$ and $^3P_2$ states via $^3S_1$.
    (b) Measured loss fraction from $^1S_0$ due to shelving as a function of probe
    detuning $\delta_p$ after cooling on the transition to $|e_\pm \rangle$ to
    $\delta_f = - 1.2\omega_g $ (diamonds) and $\delta_f = -5.5\omega_g $ (circles).
    Solid lines are numerical simulations fitted to the data to extract the ground-state fraction (see text).}
	\label{Fig3}
\end{figure}

While the above results already provide evidence that the chirp-cooling method
yields a large ground-state population, we complement the thermometry by
resolved sideband spectroscopy on the ${^1}S_0$-${^3}P_1$ transition to
$|e_0\rangle$ along the radial direction of tweezer confinement. For
spectroscopy, we apply a short ($\unit{75}{$\upmu$s}$) probe pulse with the laser frequency
set to the vicinity of the $|e_0\rangle$ resonance and the propagation direction
perpendicular to the tweezer axis. Subsequently, the population transferred to
$|e_0\rangle$ is rapidly shelved into the metastable states $^3P_0$ and $^3P_2$
(Fig.~\ref{Fig3}(a)). The sequence is repeated three times before imaging on the
${^1}S_0$ - ${^1}P_1$ transition to increase the signal. Results are shown in
Fig.~\ref{Fig3}(b) close to the lowest temperature achieved when cooling on the transition to $|e_\pm\rangle$ (diamonds). Compared to the case of less-deep cooling (circles), we observe strong sideband asymmetry, a hallmark for large ground-state population. Shift and broadening of the line with increasing temperature is due to the differential AC-Stark shift on the probe transition.
Note that without cooling, the sidebands are completely unresolved. Such effects are absent in more conventional sideband thermometry, for which
narrow optical lines with vanishing differential AC-Stark shifts are used.

Extracting the ground-state population from our data thus requires fitting with
a full numerical simulation of the spectroscopy sequence. To this end, we
compute the dynamics of the trapped two-level atom density matrix with an
initial thermal trap population as above. The population in $|e_0\rangle$ is
extracted at the end of the probe pulse and the loss fraction is determined
assuming a fitted success probability $p_s$ for shelving. Trap frequencies
$\omega_{g,e}$ and the temperature of the initial state are also fit parameters
(see Appendix \ref{SM:chirp_num}). Our lowest temperature data (diamonds) is found to be compatible
with a ground-state fraction in the range of $\unit{73}{\%}$ to $\unit{97}{\%}$, which is in agreement with the results extracted from release-and-recapture.
This ground-state fraction is also comparable with sideband cooling in magic-wavelength tweezers reported for strontium \cite{Cooper2018,Norcia2018}.

\begin{figure}[!t]
\centering
	\includegraphics[width=\columnwidth]{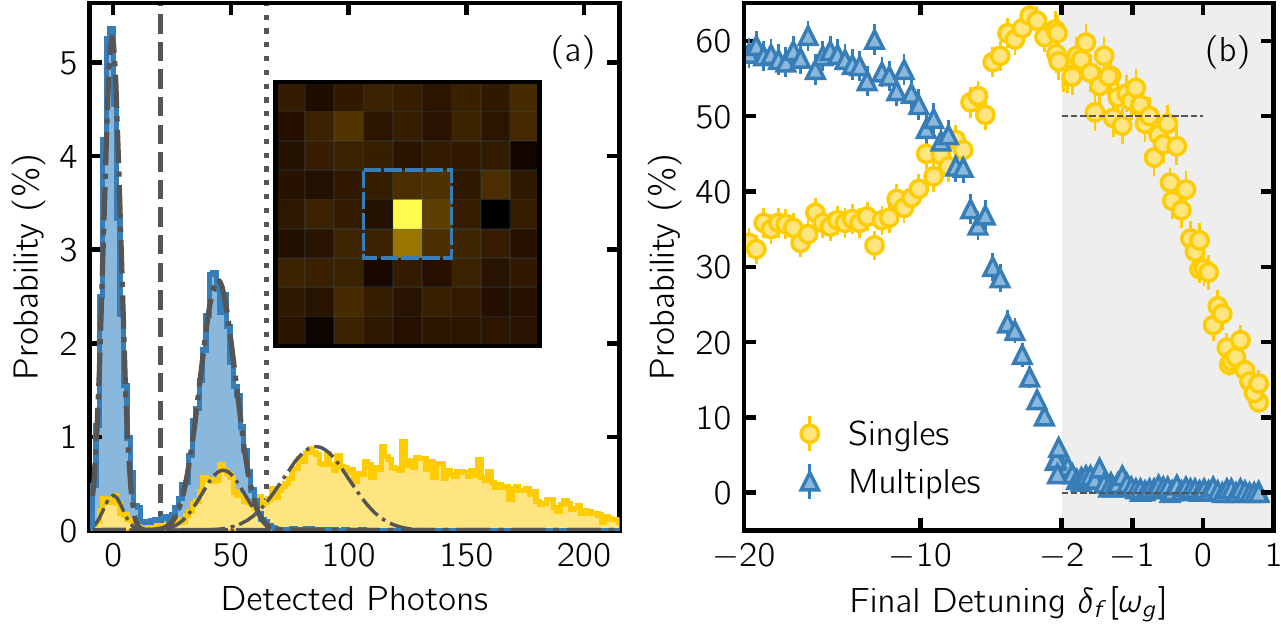}
	\caption{Light-induced losses and parity projection during cooling. (a)
  Histograms of photon counts before (yellow) and after (blue) cooling on
  $|e_\pm\rangle$  to $\delta_f =-1.2\omega_g$. The vertical dashed
  (dotted) line indicates the lower threshold set for identifying $N=1$ ($N \geq
  2$) atom(s). Dash-dotted lines are Gaussian fits to the data to guide the eye. The
  inset shows the fluorescence of a single atom imaged onto at 3x3 pixel array of
  the sCMOS camera. (b) Probability to detect $N=1$ (circles) and $N \geq 2$
  (triangles) atom(s) in the tweezer after chirp-cooling to a final detuning
  $\delta_f$. The dashed line indicates unity filling with $50 \%$ probability.
  Note the change in scaling of the axis of abscissas for $\delta_f / \omega_g > -2 $
  (shaded area).}
	\label{Fig4}
\end{figure}

Finally, we note that our data analysis neglects possible shifts
and broadening of the carrier signal due to axial temperature orthogonal to the
probe beam direction. Those can be present when probing at non-zero AC-Stark shift
due to the dependence of the radial carrier transition frequency on the axial motional
state, and allow us to infer also an upper limit estimate for the axial temperature.
Indeed, since the observed linewidth is well compatible with our 1D analysis, we conclude
that the axial temperature cannot be significantly higher than the measured radial
temperature. This provides evidence that cooling acts simultaneously in axial trap
direction, revealing additional information that is not accessible from
release-and-recapture.

\section{PARITY PROJECTION DURING COOLING}
Next, we study the dynamics of the atom number population in the tweezer
during cooling. Most importantly, we find that chirp-cooling to low temperatures
in the trap also causes light-induced losses which reliably remove pairs of
atoms from the trap \cite{Schlosser2002, Gruenzweig2010}. This can be readily
seen from histograms of the detected photon count before and after cooling
(Fig.~\ref{Fig4}(a)). Without cooling, we observe a multi-peak structure in the histogram, where the individual peaks are associated with one, two and more than two atoms in the trap. After cooling, a clean binary distribution with a one and a zero atom peak is observed. We find an approximately equal number of photon counts in the two peaks, i.e. about $50 \%$ filling with exactly one atom. In Fig.~\ref{Fig4}(b), we show how the probabilities for finding one (circles) and more than one (triangles) atom(s) in the trap evolve with the final detuning $\delta_f$ of the cooling ramp. Pairs of atoms are continuously lost with decreasing temperature and essentially vanish for $|\delta_f| \lesssim 2 \omega_g$. Indeed, this is expected as the rate for light-assisted collisions, which arise from coupling to a weakly bound molecular state below the $^1S_0$ - $^3P_1$ asymptote \cite{Cooper2018,Zelevinsky2006}, strongly depends on the wavefunction overlap of the initial pair of atoms. Moreover, along with the decreasing multi-atom signal, we find an increase in the single-atom probability, providing evidence that initial trap loading with an odd atom number $\geq 3$ results in a single atom after cooling. Thus, the chirp-cooling directly delivers parity projection, an ideal starting point for a deterministic assembly of large atom arrays \cite{Endres2016,Barredo2016}.

\section{Cooling in an array of tweezers}
Finally, we demonstrate the possibility to apply our chirp-cooling scheme to an array of multiple tweezers. To this end, we generate a one-dimensional line of ten equally spaced ($\approx \unit{10}{$\upmu$m}$) traps (Fig.~\ref{Fig5}(a)), using an acousto-optical deflector in our optical tweezer path
(see Appendix \ref{SM:tweezer_loading}). The traps are generated by applying ten RF-tones to the modulator. We equalize the trap depths to a level of $\approx \unit{2}{\%}$ via tweezer-resolved measurements of light shifts and correction on the individual RF-amplitudes. The procedure for tweezer loading and cooling is equivalent to the single-tweezer case.

\begin{figure}[!t]
\centering
	\includegraphics[width=\columnwidth]{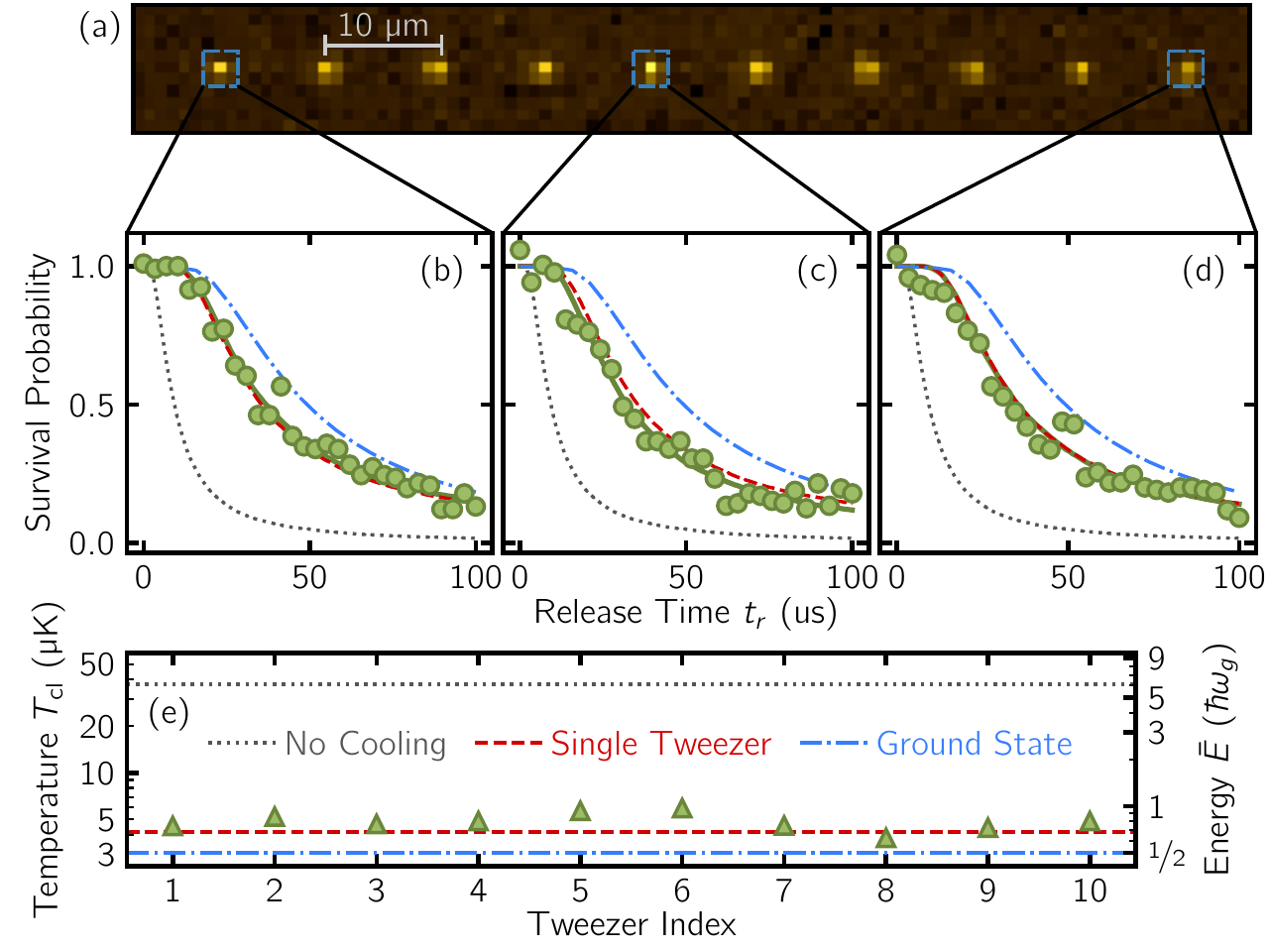}
  \caption{Chirp-cooling in a one-dimensional tweezer array. (a) Averaged fluorescence image of a line of ten tweezers (\unit{10}{$\upmu$m} spacing) after cooling chirp (including parity projection). (b)-(d) Release-and-recapture thermometry data (\textit{cf.} Fig.~\ref{Fig2}) for the three tweezers indicated with boxes after a cooling chirp to $\delta_f = - 1.2\omega_g$ on the transition to $|e_\pm\rangle$. Solid lines are fits of a classical particle trajectory simulation to extract the temperature $T_{\rm{cl}}$. The blue dash-dotted line shows the prediction of a quantum mechanical simulation for an atom in the motional ground-state. The red-dashed line (gray- dotted line) shows the classical fit to the single-tweezer data of Fig.~\ref{Fig2} for the same cooling parameters (with no cooling applied). (e) Temperatures $T_{\rm{cl}}$ as a function of the tweezer index shown in (a). Again, the red-dashed line depicts the single-tweezer result after cooling with the same cooling parameters, and the gray-dotted line is the temperature measured in a single tweezer without cooling. The blue dash-dotted line depicts the temperature-equivalent $T_{\rm{gs}}$ of the radial zero-point motion energy in the trap. Error bars show one standard deviation and are mostly smaller than the data points.}
	\label{Fig5}
\end{figure}

Exemplary release-and-recapture thermometry data for the two outermost and a central tweezer are shown in Figs. \ref{Fig5}(b)-(d) when chirp-cooling on the transition to $|e_\pm\rangle$ with $\delta_f = - 1.2\omega_g$, i.e. close to the minimally achieved temperature obtained in a single tweezer (\textit{cf.} Fig.~\ref{Fig2}). As before, the classical temperature $T_{\mathrm{cl}}$ is extracted from Monte-Carlo trajectory simulations (solid green lines), which for all ten tweezers fall almost on top of the simulation results of the single-tweezer data (red-dashed line), reported above for the same parameters. The extracted values for $T_{\mathrm{cl}}$, shown in Fig.~\ref{Fig5}(e), lie between $T_\mathrm{cl} = \unit{3.8(2)}{$\upmu$K} $ and $T_\mathrm{cl} = \unit{5.7(2)}{$\upmu$K}$, close to the single-tweezer result ($T_\mathrm{cl} = \unit{4.8(5)}{$\upmu$K}$) reported above for the same cooling parameters. The gray-dotted line again indicates the temperature without cooling. The small variations in $T_{\mathrm{cl}}$ across the array are attributed to residual differences in the trap depths. Indeed, the absolute value of $\delta_f$ required for optimal cooling depends approximately linearly on the trap depth, and in our case needs to be controlled on the percent level ($\hbar \omega_g/U_{\rm{diff}} \approx \unit{2}{\%}$).

\section{conclusion and outlook}
In conclusion, we have demonstrated a novel, broadly applicable ground-state cooling method for trapped atoms in optical tweezer arrays, which heavily releases constraints on magic trapping conditions for future experiments. This opens new routes for tweezer-based quantum technologies, requiring trapping wavelengths that have been so far incompatible with efficient in-trap cooling, specifically in view of the rapid developments with alkaline-earth(-like) atoms \cite{Madjarov2019,Barnes2022}, i.e. strontium \cite{Cooper2018,Norcia2018} or ytterbium \cite{Saskin2019}. For example, the tweezer wavelength selected in this work provides magic trapping of the two clock states $^3P_0$ and $^3P_2$, which enables new concepts for qubit encoding in a neutral atom quantum computer \cite{Pagano2022}. The cooling scheme also allows operating at wavelengths that offer additional magic trapping for Rydberg states, which mitigates decoherence and in-fidelity of two-body gates \cite{PatentFineStructureQubit}, and is also applicable to optical lattice systems \cite{Berto2021}. When simultaneously applied during imaging on the narrow intercombination transition, the reported technique may also allow for reaching higher scattering rates, particularly for atom detection in the so far unexplored case of a more strongly trapped ground state \cite{Urech2022}. Finally, we expect that applying individually controlled cooling beams along radial and axial direction in a pulsed sequence should allow for reaching the full three-dimensional trap ground-state, and leave a detailed investigation of the cooling dynamics along the weakly trapped tweezer axis for future work. In that context, it is also interesting to apply optimal control strategies on the frequency-chirped laser pulses to increase the cooling efficiency.

%%%%%%%%%%%%%%%%%%%%%%%%%%  Acknowledgments  %%%%%%%%%%%%%%%%%%%%%%%%%
\begin{acknowledgments}
We thank Johannes Zeiher, Jacob Covey, and the QRydDemo team for fruitful discussions. We acknowledge funding from the Federal Ministry of Education and Research (BMBF) under the grants CiRQus and QRydDemo, the Carl Zeiss Foundation via IQST, and the Vector Foundation. MSS acknowledges support by ONR Grant No. N00014-20-1-2513. This research was supported in part through the use of University of Delaware HPC Caviness and DARWIN computing systems.
\end{acknowledgments}

%%%%%%%%%%%%%%%%%%%%%%%%%%  Appendix  %%%%%%%%%%%%%%%%%%%%%%%%%
\appendix

\section{Tweezer loading}
\label{SM:tweezer_loading}

Tweezer loading starts with the preparation of a six-beam blue magneto-optical trap (MOT) of $^{88}$Sr atoms in a glass cell (Japan Cell), operated on the ${^1}S_0$-${^1}P_1$ ($\lambda=\unit{460.9}{nm}$) transition.
The MOT is loaded from an atomic beam source (AOSense, Inc.) comprising an oven, a Zeeman slower, and a 2D-MOT for transverse cooling.
After \unit{30}{ms} of MOT loading at a detuning of \unit{-2 \pi \times 46}{MHz} and with a saturation parameter of $s=0.12$, we decrease the detuning and saturation parameter to \unit{-2 \pi \times 21}{MHz} and $s=0.01$ within \unit{10}{ms} to reduce temperature.
The atoms are kept in this second stage of the blue MOT for another \unit{50}{ms}, before they are loaded into a narrow-line red MOT on the ${^1}S_0$-${^3}P_1$ ($\lambda=\unit{689.5}{nm}$) intercombination transition.
To this end, the magnetic field gradient is ramped from \unit{\approx 50}{G/cm} to \unit{\approx 2}{G/cm}.
During the first part of the red MOT, the cooling laser is broadened to a frequency comb with a \unit{5}{MHz} width and a regular \unit{30}{kHz} spacing by periodically modulating the RF-frequency applied to an acousto-optical modulator (AOM) to increase the capture volume.
The comb is subsequently ramped down to a single frequency with a final detuning of \unit{-2 \pi \times 150}{kHz} while simultaneously reducing the laser intensity from $s\approx 3700$ to $s\approx 45$.
After another \unit{10}{ms} hold time, the MOT contains several $10^4$ atoms at an equilibrium temperature of $\approx \unit{1.4}{$\upmu$K}$.
This atom number is found to be well-suited for loading a single tweezer and also multi-tweezer arrays, so that each site is filled with one or more atoms while also
avoiding too many atoms in a single trap. The latter may cause non-pairwise losses, ultimately leading to less than 50\% filling after parity projection \cite{Urech2022}.

For generating tweezers, we employ a frequency-doubled fiber laser system providing \unit{10}{W} output power at \unit{540}{nm} (TOPTICA Photonics). We send the trapping light through a 2D acousto-optical deflector (AA Opto-Electronic DTSXY-400) before focusing into the MOT region with a high-NA (0.5) microscope objective (Mitutoyo G Plan Apo 50X) to a waist of $\unit{564(5)}{nm}$. This allows us to extend the studies of our cooling scheme also to multi-tweezer arrays (see Fig.~\ref{Fig5}). To achieve homogeneous loading over an extended array, we increase the MOT volume by a two-step ramp of the laser detuning from \unit{- 2 \pi \times 150}{kHz} to \unit{-2 \pi \times 280}{kHz} within \unit{25}{ms} and further to \unit{-2 \pi \times 560}{kHz} within \unit{15}{ms} \cite{Cooper2018}. During the second part of this detuning ramp, the tweezer intensity is ramped up to \unit{0.85}{mW} for a single tweezer. Subsequently, the MOT beams and the magnetic quadrupole field are turned off, and the tweezer intensity is further ramped to its final value of \unit{1.7}{mW}, at which we perform our chirp-cooling experiments. Note that we illuminate the atoms with two repumping beams resonant with the ${^3}P_0$-${^3}S_1$ (\unit{679}{nm}) and ${^3}P_2$-${^3}S_1$ (\unit{707}{nm}) transitions during the entire MOT and tweezer loading procedure.

\section{Atom imaging}
\label{SM:imaging}

For single-atom detection, we induce fluorescence on the ${^1S}_0$-${^1P}_1$ transition using a separate imaging beam, pulsed on for \unit{75}{ms} with $s \approx 1.2 \times 10^{-3}$. Prior to this, the tweezer light intensity is increased to \unit{3.4}{mW}. Fluorescence photons are collected via the same objective used to focus the optical tweezers and are imaged onto a $3 \times 3$ pixel area (compare inset of Fig.~\ref{Fig4}(a) ) of a sCMOS camera (Teledyne KINETIX). The atoms are kept cold during imaging by using the red MOT beams with a fixed detuning of \unit{+ 2 \pi \times 1.2}{MHz} from the ${^1}S_0$-${^3}P_1$ free-space resonance, corresponding to a detuning from the trap-shifted transition to $|e_0\rangle$ of about \unit{- 2 \pi \times 1.4}{MHz}.

The images are classified into three categories depending on the number $N_{\rm{ph}}$ of detected photons: no atom ($N_{\rm{ph}}\leq 20$), single atom ($20 <  N_{\rm{ph}} \leq 65$), multiple atoms ($N_{\rm{ph}}>65$). We quantify the accuracy of this classification in a similar way as reported in Ref.~\cite{Cooper2018}. Taking consecutive images of the same single atom, we find a probability of $p=\unit{92.2(2)}{\%}$ to detect the atom in the second image conditioned on its detection in the first one, similar to the results reported in Refs.~\cite{Cooper2018,Norcia2018}. This atom loss during imaging also dictates the imaging fidelity. More specifically, realizations where an atom is lost from the trap result in a reduced number of detected photons, which may be below the lower threshold $N_{\rm{th}}$ set for identifying one atom. An upper bound for the probability of such false negative detection events may be estimated via $1-{\rm{exp}}({\rm{ln}}(p) N_{\rm{th}} / \bar{N})$, where $\bar{N}$ is the mean number of detected photons for one atom \cite{Cooper2018}. For the imaging parameters used throughout this work, we find a false negative rate of about \unit{3.6(1)}{\%}. The false positive rate for identifying a single atom (\unit{0.00(2)}{\%}) is negligible.

To extract the survival probability in Fig.~\ref{Fig2}(a) and the loss fraction in Fig.~\ref{Fig3}(b), a value $\zeta_c$ is assigned to each of the imaging classification outcomes: $\zeta_c=0$ (no atom), $\zeta_c=1$ (single atom), and $\zeta_c=2$ for multiple atoms, which is then averaged for each data point. To improve the signal-to-noise ratio of the sideband spectroscopy data in Fig.~\ref{Fig3}(c), we take two consecutive images. The first one is taken after the shelving spectroscopy pulse. Before taking the second image, we return the shelved population from the long-lived $^3P_0$ and $^3P_2$ states back into the imaging cycle by applying the \unit{679}{nm} and \unit{707}{nm} repumping lasers. We then post-select on realizations where an atom is detected in the second image.

Finally, we note that the accuracy to distinguish between single and multiple atoms in the trap is much lower than between zero and one atom, since the corresponding signals in the histograms (Fig.~\ref{Fig4}(a)) strongly overlap.
The characterization threshold for multiple atoms is chosen high enough that it does not affect the low-temperature measurements with a clean bimodal distribution, i.e. after successful parity projection.

\section{Differential AC-Stark shifts}
\label{SM:diffac}

In Fig.~\ref{Fig1}(d), we plot the ratio of the AC-polarizabilities (and hence the trap
depths) between the $^1S_0$ electronic ground-state and the trap-shifted sublevels $|e_0 \rangle$
and $|e_\pm \rangle$ of the $^3P_1$ excited state. For our case of nominally zero magnetic field and
a linearly polarized tweezer, the AC-Stark shift for $^3P_1$ can be readily expressed in terms of scalar ($\alpha_s$) and tensor $(\alpha_t)$ polarizabilities (the contribution of the vector polarizability vanishes in the absence of magnetic field and for linear polarization \cite{Kien2013,Cooper2018}). The polarizability of the $^1S_0$ ground-state has only a scalar contribution.

More specifically, we consider the time-independent AC-Stark interaction Hamiltonian with an optical field $\vec{E}(t)=\vec{E}^+ e^{-i \omega t}+ \vec{E}^- e^{+i \omega t}$, where $\vec{E}^+ = E_0 \vec{\epsilon}$, $\vec{\epsilon}$
being the polarization vector, and $\vec{E}^-$ the complex conjugate of $\vec{E}^+$. Omitting the vector term, the Hamiltonian reads
\begin{equation}
\begin{split}
\mathcal{H} & =  -\alpha_s E_0^2 \\
& - \frac{3 \alpha_t}{J(2J-1)} \left( \frac{\{\vec{E}^+ \cdot \vec{J}, \vec{E}^- \cdot \vec{J} \}}{2} - \frac{J(J+1) E_0^2}{3} \right) .
\end{split}
\end{equation}
Here, $\vec{J} = (J_x,J_y,J_z)$ denotes the total angular momentum operator and $J$ the associated
angular momentum quantum number. In the absence of an external magnetic field, it is convenient to
define the quantization axis along the tweezer polarization, which we set (without loss of
generality) along the $x$-direction. Accordingly, we label the bare Zeeman substates of the $^3P_1$
level as $|m_J = 0, \pm 1\rangle$, where $m_J$ is the magnetic quantum number associated with the
projection of the total angular momentum along the tweezer polarization. The AC-Stark Hamiltonian then reduces to
\begin{equation}
\mathcal{H_{\rm{lin}}} =  - E_0^2 \left( \alpha_s +  \alpha_t \frac{3 J_x^2 - J(J+1)}{2(2J-1)} \right) .
\end{equation}
Since the Hamiltonian above is diagonal in the $|m_J = 0, \pm 1\rangle$ basis defined along the $x$-direction, the total polarizability reads \cite{Cooper2018}
\begin{equation}
\alpha = \alpha_s + \alpha_t \frac{3 m_J^2 - J(J+1)}{J(2J-1)} \, ,
\end{equation}
For $J=1$ ($^3P_1$) one finds $\alpha = \alpha_s - 2 \alpha_t$ for $|e_0\rangle = |m_J = 0\rangle$ and $\alpha = \alpha_s + \alpha_t$ for $|e_\pm \rangle = |m_J = \pm 1\rangle$. Wavelength-dependent values for $\alpha_s$ and $\alpha_t$ are obtained from numerical calculations as follows.

We evaluated the dynamic polarizabilities by solving the inhomogeneous equation in valence space \cite{Kozlov1999} using the Dalgarno-Lewis \cite{Dalgarno1955} approach. This approach allows to account for both discrete states and the continuum. We find intermediate-state wave functions $\delta \psi_{\pm}$ from an inhomogeneous equation,
\begin{eqnarray}
|\delta \psi_{\pm} \rangle & = & \frac{1}{H_{\rm eff} - E_0 \pm \omega}\,
 \sum_k | \Psi_k \rangle \langle \Psi_k | D | \Psi_0 \rangle \nonumber \\
&=&  \frac{1}{H_{\rm eff}- E_0 \pm \omega} \, D| \Psi_0 \rangle,
\label{delpsi}
\end{eqnarray}
where $D$ is the $z$-component of the effective electric dipole operator ${\bf D}$,  $\Psi_0$ is the wave function, and $E_0$ is the energy of the state of interest, either $^1S_0$ or $^3P_1$ in the present work.

The wave functions are computed using the relativistic high-precision hybrid method that combines configuration interaction and coupled-cluster approaches (CI+all-order) \cite{Cheung2021,Safronova2009}.
In this method, the energies and wave functions are determined from the time-independent multiparticle Schr\"odinger equation
\begin{equation}
H_{\rm eff}(E_k) \Psi_k = E_k \Psi_k,
\label{Heff}
\end{equation}
where the effective Hamiltonian  $H_{\rm eff}$ includes contributions of the core states constructed using the coupled-cluster method.

The  polarizability  is  given by
\begin{equation}
\alpha_v (\omega ) = \langle \Psi_0 |D_0| \delta \psi_+ \rangle
+ \langle \Psi_0 |D_0| \delta \psi_- \rangle  \, ,
\label{alpha2}
\end{equation}
where $v$ indicates that this method gives the valence contribution to the polarizability.
The small core polarizability contribution is computed in the random-phase approximation.
One of the challenges of the accurate polarizability computation for the  $^3P_1$ state in the region below \unit{600}{nm} is strong sensitivity to the accuracy of the  energy levels. To resolve this problem, we developed a code to automatically replace the theoretical energy values for low-lying dominant contributions by exact experimental values as well as use improved recommended values of the reduced matrix element where available. The substitution is done for all data points using the sum-over-states formula
\begin{eqnarray}
\alpha_v (\omega) &=& 2\, \sum_k \frac { \left( E_k-E_0 \right)
|\langle \Psi_0 |D_0| \Psi_k \rangle|^2 }
      { \left( E_k-E_0 \right)^2 - \omega^2 },  \nonumber \\
\end{eqnarray}
improving the polarizability accuracy. The uncertainties are estimated for all polarizability values.

\section{Classical and quantum mechanical release-and-recapture analysis}
\label{SM:rr_num}

In this section, we provide details on the classical and quantum mechanical analysis of the release-and-recapture data shown in Fig.~\ref{Fig2}. Our classical analysis follows the procedure described in Ref.~\cite{Tuchendler2008}. Specifically, we draw a Monte-Carlo sample of spatial and velocity vectors from a thermal distribution of point particles in a 3D harmonic trap using our experimental parameters. For a given classical temperature $T_{\rm{cl}}$, the three spatial coordinates are Gauss-distributed with a standard deviation of $\sigma_i = \sqrt{k_B T_{\rm{cl}} / m \omega_i^2}$, where $m$ is the mass of the $^{88}$Sr atom, and $\omega_i$ the oscillator trap frequency in direction $i=(x,y,z)$. The Gaussian velocity distribution in each direction has a standard deviation of $\sigma_v = \sqrt{k_B T_{\rm{cl}} / m}$. A simulated release-and-recapture trace is obtained by propagating this ensemble in free-space for a variable time $t_r$. The survival probability is then computed by evaluating the fraction of the ensemble that is trapped after instantaneous switch-on of the Gaussian tweezer potential. A particle is considered to be trapped, when its kinetic energy is smaller than the local (absolute) potential energy after the free propagation. Such simulated traces are then fit to the data via a chi-squared analysis with $T_{\rm{cl}}$ as fit parameter.

This analysis does not capture the zero-point motion energy of the trapped atom, and the fitted classical temperature $T_{\rm{cl}}$ overestimates the true quantum mechanical temperature. Effects of quantized motion in the trap are taken into account by our quantum mechanical analysis of the lowest energy data in Fig.~\ref{Fig2}(a). To this end, it is sufficient to consider only the radial dynamics, since the release-and-recapture technique is only weakly dependent on the longitudinal motion. First, we compute the free expansion of initial 2D harmonic oscillator wavefunctions with occupation numbers $(\nu_x,\nu_y)$, where $x$ and $y$ denote the cartesian coordinates of the radial tweezer direction. After a variable time $t_r$, the Gaussian tweezer potential is turned on instantaneously and the wavefunction is evolved inside the potential for another \unit{100}{$\upmu$s}. The fraction of the wavefunction that has remained inside the potential then yields the survival probability $p_{(\nu_x,\nu_y)}$. Assuming a thermal population of the 2D harmonic oscillator levels, the survival probability $p_{\rm{qm}}$ for an ensemble at a quantum mechanical temperature $T_{\rm{qm}}$ is finally obtained by weighting $p_{(\nu_x,\nu_y)}$ with the corresponding Boltzmann factor,
\begin{equation}
p_{\rm{qm}} = \frac{1}{Z} \, p_{(\nu_x,\nu_y)} \, e^{- \hbar \omega_{(x,y)} (\nu_x + \nu_y +1 ) / (k_B T_{\rm{qm}})}.
\end{equation}
Here, $\omega_{(x,y)}$ denotes the radial trap frequency and $Z$ the partition function of the 2D harmonic oscillator. We fit the simulated $p_{\rm{qm}}$ to the data with $T_{\rm{qm}}$ as fit parameter. Finally, the fitted value for $T_{\rm{qm}}$ yields the ground-state population along one radial direction stated in the main article.

\section{Numerical modeling of chirp-cooling and sideband spectroscopy data}
\label{SM:chirp_num}

In this section, we provide details of our method for describing the chirp-cooling and for modeling the sideband spectroscopy data. Our method is similar to the approach taken in \cite{Berto2021,Taieb1994}. To reduce the computational costs of our simulations, we neglect coupling between different spatial directions and restrict ourselves to one radial direction of the three-dimensional trap. Under this approximation, the system can be described as a driven two-level atom in a one-dimensional harmonic trap. The electronic ground-state $\ket{g}$ is trapped with the frequency $\omega_g$ and the excited state $\ket{e}$ with $\omega_e$. We use the levels of the harmonic oscillator with the frequency $\omega_g$ as a basis for the motional state of the atom. For our simulations, we take the 30 lowest oscillator levels into account. Using the creation and annihilation operators $a^\dagger$ and $a$ that act on these levels, the Hamiltonian of the system reads \cite{Kale2020,Taieb1994}
\begin{equation}
H = \hbar\omega_g \left(a^\dag a + \frac{1}{2}\right) + H_\text{atom}(t) + H_\text{int} \,.
\end{equation}

To derive the electronic Hamiltonian $H_\text{atom}(t)$ of the laser-driven atom, we change into the rotating frame of the laser. Using the rotating wave approximation, we obtain
\begin{equation}
\begin{split}
H_\text{atom}(t) = &
\frac{\hbar\Omega}{2} e^{i \eta (a^\dag + a)} \ket{e}\bra{g} +
\frac{\hbar\Omega^*}{2} e^{-i \eta (a^\dag + a)} \ket{g}\bra{e} \\
& -\hbar \delta(t) \ket{e}\bra{e} \, ,
\end{split}
\end{equation}
with the time-dependent detuning $\delta(t)$, Rabi frequency $\Omega$, Lamb-Dicke parameter $\eta = k x_0 = \tfrac{2\pi}{\lambda} \sqrt{\tfrac{\hbar}{2 m \omega}}$, wavelength $\lambda=689 \;\text{nm}$, and $m$ being the mass of the $^{88}$Sr atom.

\begin{figure}[!t]
\centering
	\includegraphics[width=\columnwidth]{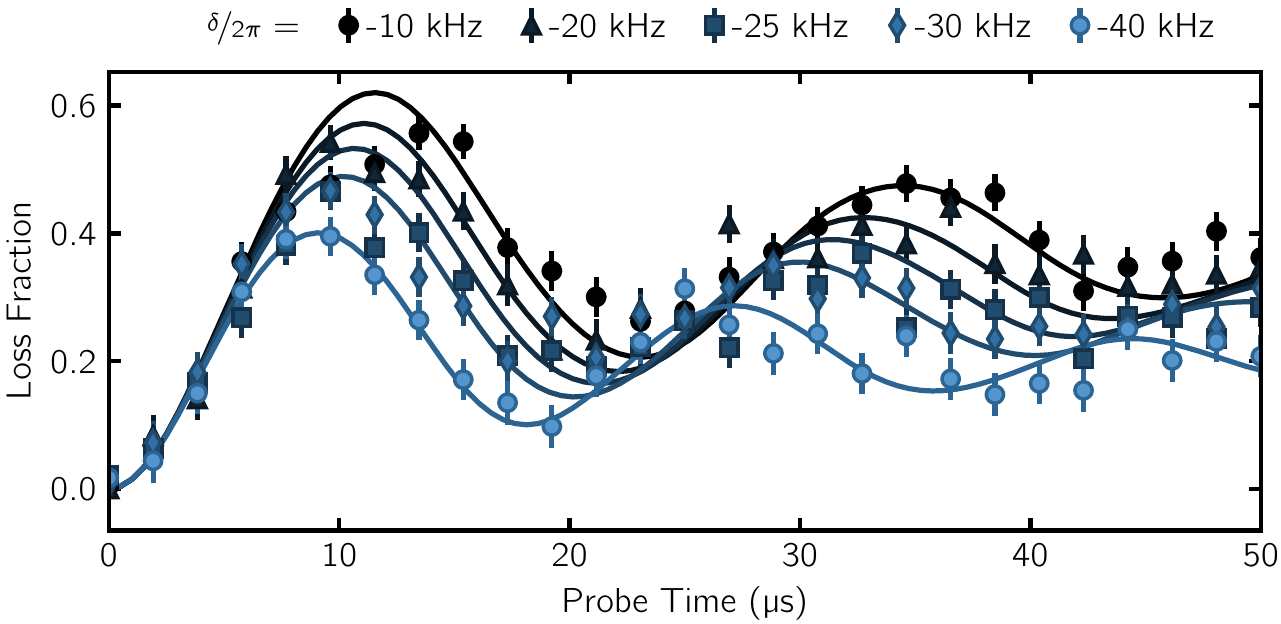}
  \caption{Measurement of the probe Rabi frequency $\Omega$ used for sideband spectroscopy. Measured loss fraction from ${^1}S_0$ due to shelving as a function of probe time for different detunings from the carrier transition, i.e. $\nu_e = \nu_g$, to $|e_0\rangle$. Solid lines show numerical simulations fit to the data, which yield $\Omega = 2 \pi \times \unit{44.9(5)}{kHz}$.}
	\label{Figsup1}
\end{figure}

The interaction $H_\text{int}$ between the electronic and motional states of the atom emerges from the trap frequency being state-dependent and is given by
\begin{equation}
H_\text{int} =
\frac{\hbar(\omega_e^2-\omega_g^2)}{4\omega_g} \; \left(a+a^\dag\right)^2  \; \ket{e}\bra{e} +
U \ket{e}\bra{e}\,,
\end{equation}
where $U$ is chosen such that the transition from $\ket{g}$ to $\ket{e}$ is driven resonantly for $\delta(t) = 0$ if the atom is in its motional ground-state.

To incorporate the decay of the excited state with rate $\gamma=2\pi \times 7.4 \;\text{kHz}$, we describe the system with a Lindblad master equation, $\dot{\rho} = -i(H_\text{eff} \rho - \rho H_\text{eff}) + \mathcal{L} \rho$. The decay enters the non-hermitian Hamiltonian $H_\text{eff}=H-\tfrac{i\hbar\gamma}{2} \ket{e} \bra{e}$ and the term
\begin{equation}
\begin{split}
\mathcal{L}\rho & =  \frac{\hbar\gamma}{2} \int_0^\pi  e^{-i \eta (a^\dag + a) \cos \theta} \ket{g}\bra{e} \rho \ket{e} \bra{g} \,  \\
                & \times\;  e^{i \eta (a^\dag + a) \cos \theta} \sin \theta\; \mathrm{d}\theta \,.
\end{split}
\end{equation}
This term accounts for the population returning into the ground-state and for the recoil of the emitted photon projected onto the direction of the trap.

To illustrate the chirp-cooling approach (Fig.~\ref{Fig1}(b)), we apply a Rabi frequency of $\Omega=2\pi \times 20 \,\text{kHz}$ and a time-dependent detuning $\delta(t)$ that is ramped linearly from $-3.7\, \omega_g$ to $-1\; \omega_g$ within $10 \,\text{ms}$. The trap frequencies are set to $\omega_g=2\pi \times 150 \,\text{kHz}$ and $\omega_e=2\pi \times 110 \;\text{kHz}$. To demonstrate the cooling mechanism, we use the motionally excited state $\ket{g,4}$ as  the initial state. The resulting Lindblad master equation is solved using QuTiP \cite{Johansson2013}.

For modeling the sideband spectroscopy data, we extend our model by introducing an effective dark state. The sideband spectroscopy is a two-step process. First, we apply a probe pulse with Rabi frequency $\Omega$ and detuning $\delta$ for $\unit{75}{$\upmu$s}$. Second, we transfer the resulting population in $\ket{e}$ to the dark state with a success probability $p_s$. The probe pulse is again simulated using QuTiP, using a thermal density matrix with temperature $T$ as initial state. In the experiment, the shelving signal is increased by repeating the spectroscopy sequence three times before imaging. To account for this in our simulation, we also compute the pulse sequence three times, and take the populations of oscillator states after each pulse (with the atom measured in $\ket{g}$) as the initial condition for the next pulse. In doing so, we include the influence of the probe pulse on the population of motional states, which is experimentally relevant since the lifetime of $\ket{e}$ ($1/\gamma = \unit{21.5}{$\upmu$s} $) is comparable to the probe pulse length. Repeating this procedure for various values of $\delta$ yields the simulated sideband spectra. Finally, we fit the simulation results to the experimentally measured data with free parameters $T$, $\omega_g$, and $p_s$. The Rabi frequency $\Omega = 2 \pi \times \unit{44.9(5)}{kHz}$ is measured independently (see Fig.~\ref{Figsup1} and text below), and the excited state trap frequency $\omega_e$ is computed from the calculated polarizability ratio of the transition to $|e_0\rangle$. The fit yields $\omega_g=2 \pi \times \unit{126(5)}{kHz}$, i.e. the trap frequency stated in the main article and $p_s = 0.33$. The range of ground-state fraction reported in the main text reflects the set of simulated sideband spectra when varying $T$ that are compatible with the data within the experimental error bars.

Finally, we discuss briefly the independent measurement of $\Omega$, for which we apply our sideband spectroscopy sequence as before but now vary
the length of the probe pulses. Measured shelving data as a function
of probe time exhibits damped Rabi oscillations (see Fig.~\ref{Figsup1}). To extract $\Omega$, we fit the data with the same numerical simulations as described above,
but now vary the time of the probe pulses for different values of $\delta$.

% \bibliography{lit}

\begin{thebibliography}{41}%
\makeatletter
\providecommand \@ifxundefined [1]{%
 \@ifx{#1\undefined}
}%
\providecommand \@ifnum [1]{%
 \ifnum #1\expandafter \@firstoftwo
 \else \expandafter \@secondoftwo
 \fi
}%
\providecommand \@ifx [1]{%
 \ifx #1\expandafter \@firstoftwo
 \else \expandafter \@secondoftwo
 \fi
}%
\providecommand \natexlab [1]{#1}%
\providecommand \enquote  [1]{``#1''}%
\providecommand \bibnamefont  [1]{#1}%
\providecommand \bibfnamefont [1]{#1}%
\providecommand \citenamefont [1]{#1}%
\providecommand \href@noop [0]{\@secondoftwo}%
\providecommand \href [0]{\begingroup \@sanitize@url \@href}%
\providecommand \@href[1]{\@@startlink{#1}\@@href}%
\providecommand \@@href[1]{\endgroup#1\@@endlink}%
\providecommand \@sanitize@url [0]{\catcode `\\12\catcode `\$12\catcode `\&12\catcode `\#12\catcode `\^12\catcode `\_12\catcode `\%12\relax}%
\providecommand \@@startlink[1]{}%
\providecommand \@@endlink[0]{}%
\providecommand \url  [0]{\begingroup\@sanitize@url \@url }%
\providecommand \@url [1]{\endgroup\@href {#1}{\urlprefix }}%
\providecommand \urlprefix  [0]{URL }%
\providecommand \Eprint [0]{\href }%
\providecommand \doibase [0]{https://doi.org/}%
\providecommand \selectlanguage [0]{\@gobble}%
\providecommand \bibinfo  [0]{\@secondoftwo}%
\providecommand \bibfield  [0]{\@secondoftwo}%
\providecommand \translation [1]{[#1]}%
\providecommand \BibitemOpen [0]{}%
\providecommand \bibitemStop [0]{}%
\providecommand \bibitemNoStop [0]{.\EOS\space}%
\providecommand \EOS [0]{\spacefactor3000\relax}%
\providecommand \BibitemShut  [1]{\csname bibitem#1\endcsname}%
\let\auto@bib@innerbib\@empty
%</preamble>
\bibitem [{\citenamefont {Kaufman}\ and\ \citenamefont {Ni}(2021)}]{Kaufman2021}%
  \BibitemOpen
  \bibfield  {author} {\bibinfo {author} {\bibfnamefont {A.~M.}\ \bibnamefont {Kaufman}}\ and\ \bibinfo {author} {\bibfnamefont {K.-K.}\ \bibnamefont {Ni}},\ }\bibfield  {title} {\bibinfo {title} {Quantum science with optical tweezer arrays of ultracold atoms and molecules},\ }\href{https://www.nature.com/articles/s41567-021-01357-2} {\bibfield  {journal} {\bibinfo  {journal} {Nature Physics}\ }\textbf {\bibinfo {volume} {17}},\ \bibinfo {pages} {1324} (\bibinfo {year} {2021})}\BibitemShut {NoStop}%
\bibitem [{\citenamefont {Browaeys}\ and\ \citenamefont {Lahaye}(2020)}]{Browaeys2020}%
  \BibitemOpen
  \bibfield  {author} {\bibinfo {author} {\bibfnamefont {A.}~\bibnamefont {Browaeys}}\ and\ \bibinfo {author} {\bibfnamefont {T.}~\bibnamefont {Lahaye}},\ }\bibfield  {title} {\bibinfo {title} {Many-body physics with individually controlled rydberg atoms},\ }\href {https://doi.org/10.1038/s41567-019-0733-z} {\bibfield  {journal} {\bibinfo  {journal} {Nature Physics}\ }\textbf {\bibinfo {volume} {16}},\ \bibinfo {pages} {132} (\bibinfo {year} {2020})}\BibitemShut {NoStop}%
\bibitem [{\citenamefont {Bluvstein}\ \emph {et~al.}(2022)\citenamefont {Bluvstein}, \citenamefont {Levine}, \citenamefont {Semeghini}, \citenamefont {Wang}, \citenamefont {Ebadi}, \citenamefont {Kalinowski}, \citenamefont {Keesling}, \citenamefont {Maskara}, \citenamefont {Pichler}, \citenamefont {Greiner}, \citenamefont {Vuleti{\'{c}}},\ and\ \citenamefont {Lukin}}]{Bluvstein2022}%
  \BibitemOpen
  \bibfield  {author} {\bibinfo {author} {\bibfnamefont {D.}~\bibnamefont {Bluvstein}}, \bibinfo {author} {\bibfnamefont {H.}~\bibnamefont {Levine}}, \bibinfo {author} {\bibfnamefont {G.}~\bibnamefont {Semeghini}}, \bibinfo {author} {\bibfnamefont {T.~T.}\ \bibnamefont {Wang}}, \bibinfo {author} {\bibfnamefont {S.}~\bibnamefont {Ebadi}}, \bibinfo {author} {\bibfnamefont {M.}~\bibnamefont {Kalinowski}}, \bibinfo {author} {\bibfnamefont {A.}~\bibnamefont {Keesling}}, \bibinfo {author} {\bibfnamefont {N.}~\bibnamefont {Maskara}}, \bibinfo {author} {\bibfnamefont {H.}~\bibnamefont {Pichler}}, \bibinfo {author} {\bibfnamefont {M.}~\bibnamefont {Greiner}}, \bibinfo {author} {\bibfnamefont {V.}~\bibnamefont {Vuleti{\'{c}}}},\ and\ \bibinfo {author} {\bibfnamefont {M.~D.}\ \bibnamefont {Lukin}},\ }\bibfield  {title} {\bibinfo {title} {A quantum processor based on coherent transport of entangled atom arrays},\ }\href {https://doi.org/10.1038/s41586-022-04592-6} {\bibfield  {journal} {\bibinfo  {journal} {Nature}\ }\textbf {\bibinfo {volume} {604}},\ \bibinfo {pages} {451} (\bibinfo {year} {2022})}\BibitemShut {NoStop}%
\bibitem [{\citenamefont {Graham}\ \emph {et~al.}(2022)\citenamefont {Graham}, \citenamefont {Song}, \citenamefont {Scott}, \citenamefont {Poole}, \citenamefont {Phuttitarn}, \citenamefont {Jooya}, \citenamefont {Eichler}, \citenamefont {Jiang}, \citenamefont {Marra}, \citenamefont {Grinkemeyer}, \citenamefont {Kwon}, \citenamefont {Ebert}, \citenamefont {Cherek}, \citenamefont {Lichtman}, \citenamefont {Gillette}, \citenamefont {Gilbert}, \citenamefont {Bowman}, \citenamefont {Ballance}, \citenamefont {Campbell}, \citenamefont {Dahl}, \citenamefont {Crawford}, \citenamefont {Blunt}, \citenamefont {Rogers}, \citenamefont {Noel},\ and\ \citenamefont {Saffman}}]{Graham2022}%
  \BibitemOpen
  \bibfield  {author} {\bibinfo {author} {\bibfnamefont {T.~M.}\ \bibnamefont {Graham}}, \bibinfo {author} {\bibfnamefont {Y.}~\bibnamefont {Song}}, \bibinfo {author} {\bibfnamefont {J.}~\bibnamefont {Scott}}, \bibinfo {author} {\bibfnamefont {C.}~\bibnamefont {Poole}}, \bibinfo {author} {\bibfnamefont {L.}~\bibnamefont {Phuttitarn}}, \bibinfo {author} {\bibfnamefont {K.}~\bibnamefont {Jooya}}, \bibinfo {author} {\bibfnamefont {P.}~\bibnamefont {Eichler}}, \bibinfo {author} {\bibfnamefont {X.}~\bibnamefont {Jiang}}, \bibinfo {author} {\bibfnamefont {A.}~\bibnamefont {Marra}}, \bibinfo {author} {\bibfnamefont {B.}~\bibnamefont {Grinkemeyer}}, \bibinfo {author} {\bibfnamefont {M.}~\bibnamefont {Kwon}}, \bibinfo {author} {\bibfnamefont {M.}~\bibnamefont {Ebert}}, \bibinfo {author} {\bibfnamefont {J.}~\bibnamefont {Cherek}}, \bibinfo {author} {\bibfnamefont {M.~T.}\ \bibnamefont {Lichtman}}, \bibinfo {author} {\bibfnamefont {M.}~\bibnamefont {Gillette}}, \bibinfo {author} {\bibfnamefont {J.}~\bibnamefont {Gilbert}}, \bibinfo {author} {\bibfnamefont {D.}~\bibnamefont {Bowman}}, \bibinfo {author} {\bibfnamefont {T.}~\bibnamefont {Ballance}}, \bibinfo {author} {\bibfnamefont {C.}~\bibnamefont {Campbell}}, \bibinfo {author} {\bibfnamefont {E.~D.}\ \bibnamefont {Dahl}}, \bibinfo {author} {\bibfnamefont {O.}~\bibnamefont {Crawford}}, \bibinfo {author} {\bibfnamefont {N.~S.}\ \bibnamefont {Blunt}}, \bibinfo {author} {\bibfnamefont {B.}~\bibnamefont {Rogers}}, \bibinfo {author} {\bibfnamefont {T.}~\bibnamefont {Noel}},\ and\ \bibinfo {author} {\bibfnamefont {M.}~\bibnamefont {Saffman}},\ }\bibfield  {title} {\bibinfo {title} {Multi-qubit entanglement and algorithms on a neutral-atom quantum computer},\ }\href {https://doi.org/10.1038/s41586-022-04603-6} {\bibfield  {journal} {\bibinfo  {journal} {Nature}\ }\textbf {\bibinfo {volume} {604}},\ \bibinfo {pages} {457} (\bibinfo {year} {2022})}\BibitemShut {NoStop}%
\bibitem [{\citenamefont {Madjarov}\ \emph {et~al.}(2019)\citenamefont {Madjarov}, \citenamefont {Cooper}, \citenamefont {Shaw}, \citenamefont {Covey}, \citenamefont {Schkolnik}, \citenamefont {Yoon}, \citenamefont {Williams},\ and\ \citenamefont {Endres}}]{Madjarov2019}%
  \BibitemOpen
  \bibfield  {author} {\bibinfo {author} {\bibfnamefont {I.~S.}\ \bibnamefont {Madjarov}}, \bibinfo {author} {\bibfnamefont {A.}~\bibnamefont {Cooper}}, \bibinfo {author} {\bibfnamefont {A.~L.}\ \bibnamefont {Shaw}}, \bibinfo {author} {\bibfnamefont {J.~P.}\ \bibnamefont {Covey}}, \bibinfo {author} {\bibfnamefont {V.}~\bibnamefont {Schkolnik}}, \bibinfo {author} {\bibfnamefont {T.~H.}\ \bibnamefont {Yoon}}, \bibinfo {author} {\bibfnamefont {J.~R.}\ \bibnamefont {Williams}},\ and\ \bibinfo {author} {\bibfnamefont {M.}~\bibnamefont {Endres}},\ }\bibfield  {title} {\bibinfo {title} {An atomic-array optical clock with single-atom readout},\ }\href {https://doi.org/10.1103/physrevx.9.041052} {\bibfield  {journal} {\bibinfo  {journal} {Physical Review X}\ }\textbf {\bibinfo {volume} {9}},\ \bibinfo {pages} {041052} (\bibinfo {year} {2019})}\BibitemShut {NoStop}%
\bibitem [{\citenamefont {Young}\ \emph {et~al.}(2020)\citenamefont {Young}, \citenamefont {Eckner}, \citenamefont {Milner}, \citenamefont {Kedar}, \citenamefont {Norcia}, \citenamefont {Oelker}, \citenamefont {Schine}, \citenamefont {Ye},\ and\ \citenamefont {Kaufman}}]{Young2020}%
  \BibitemOpen
  \bibfield  {author} {\bibinfo {author} {\bibfnamefont {A.~W.}\ \bibnamefont {Young}}, \bibinfo {author} {\bibfnamefont {W.~J.}\ \bibnamefont {Eckner}}, \bibinfo {author} {\bibfnamefont {W.~R.}\ \bibnamefont {Milner}}, \bibinfo {author} {\bibfnamefont {D.}~\bibnamefont {Kedar}}, \bibinfo {author} {\bibfnamefont {M.~A.}\ \bibnamefont {Norcia}}, \bibinfo {author} {\bibfnamefont {E.}~\bibnamefont {Oelker}}, \bibinfo {author} {\bibfnamefont {N.}~\bibnamefont {Schine}}, \bibinfo {author} {\bibfnamefont {J.}~\bibnamefont {Ye}},\ and\ \bibinfo {author} {\bibfnamefont {A.~M.}\ \bibnamefont {Kaufman}},\ }\bibfield  {title} {\bibinfo {title} {Half-minute-scale atomic coherence and high relative stability in a tweezer clock},\ }\href {https://doi.org/10.1038/s41586-020-3009-y} {\bibfield  {journal} {\bibinfo  {journal} {Nature}\ }\textbf {\bibinfo {volume} {588}},\ \bibinfo {pages} {408} (\bibinfo {year} {2020})}\BibitemShut {NoStop}%
\bibitem [{\citenamefont {Cheuk}\ \emph {et~al.}(2020)\citenamefont {Cheuk}, \citenamefont {Anderegg}, \citenamefont {Bao}, \citenamefont {Burchesky}, \citenamefont {Yu}, \citenamefont {Ketterle}, \citenamefont {Ni},\ and\ \citenamefont {Doyle}}]{Cheuk2020}%
  \BibitemOpen
  \bibfield  {author} {\bibinfo {author} {\bibfnamefont {L.~W.}\ \bibnamefont {Cheuk}}, \bibinfo {author} {\bibfnamefont {L.}~\bibnamefont {Anderegg}}, \bibinfo {author} {\bibfnamefont {Y.}~\bibnamefont {Bao}}, \bibinfo {author} {\bibfnamefont {S.}~\bibnamefont {Burchesky}}, \bibinfo {author} {\bibfnamefont {S.~S.}\ \bibnamefont {Yu}}, \bibinfo {author} {\bibfnamefont {W.}~\bibnamefont {Ketterle}}, \bibinfo {author} {\bibfnamefont {K.-K.}\ \bibnamefont {Ni}},\ and\ \bibinfo {author} {\bibfnamefont {J.~M.}\ \bibnamefont {Doyle}},\ }\bibfield  {title} {\bibinfo {title} {Observation of collisions between two ultracold ground-state {CaF} molecules},\ }\href {https://doi.org/10.1103/physrevlett.125.043401} {\bibfield  {journal} {\bibinfo  {journal} {Physical Review Letters}\ }\textbf {\bibinfo {volume} {125}},\ \bibinfo {pages} {043401} (\bibinfo {year} {2020})}\BibitemShut {NoStop}%
\bibitem [{\citenamefont {Cairncross}\ \emph {et~al.}(2021)\citenamefont {Cairncross}, \citenamefont {Zhang}, \citenamefont {Picard}, \citenamefont {Yu}, \citenamefont {Wang},\ and\ \citenamefont {Ni}}]{Cairncross2021}%
  \BibitemOpen
  \bibfield  {author} {\bibinfo {author} {\bibfnamefont {W.~B.}\ \bibnamefont {Cairncross}}, \bibinfo {author} {\bibfnamefont {J.~T.}\ \bibnamefont {Zhang}}, \bibinfo {author} {\bibfnamefont {L.~R.}\ \bibnamefont {Picard}}, \bibinfo {author} {\bibfnamefont {Y.}~\bibnamefont {Yu}}, \bibinfo {author} {\bibfnamefont {K.}~\bibnamefont {Wang}},\ and\ \bibinfo {author} {\bibfnamefont {K.-K.}\ \bibnamefont {Ni}},\ }\bibfield  {title} {\bibinfo {title} {Assembly of a rovibrational ground state molecule in an optical tweezer},\ }\href {https://doi.org/10.1103/physrevlett.126.123402} {\bibfield  {journal} {\bibinfo  {journal} {Physical Review Letters}\ }\textbf {\bibinfo {volume} {126}},\ \bibinfo {pages} {123402} (\bibinfo {year} {2021})}\BibitemShut {NoStop}%
\bibitem [{\citenamefont {Thompson}\ \emph {et~al.}(2013)\citenamefont {Thompson}, \citenamefont {Tiecke}, \citenamefont {Zibrov}, \citenamefont {Vuleti{\'{c}}},\ and\ \citenamefont {Lukin}}]{Thompson2013}%
  \BibitemOpen
  \bibfield  {author} {\bibinfo {author} {\bibfnamefont {J.~D.}\ \bibnamefont {Thompson}}, \bibinfo {author} {\bibfnamefont {T.~G.}\ \bibnamefont {Tiecke}}, \bibinfo {author} {\bibfnamefont {A.~S.}\ \bibnamefont {Zibrov}}, \bibinfo {author} {\bibfnamefont {V.}~\bibnamefont {Vuleti{\'{c}}}},\ and\ \bibinfo {author} {\bibfnamefont {M.~D.}\ \bibnamefont {Lukin}},\ }\bibfield  {title} {\bibinfo {title} {Coherence and raman sideband cooling of a single atom in an optical tweezer},\ }\href {https://doi.org/10.1103/physrevlett.110.133001} {\bibfield  {journal} {\bibinfo  {journal} {Physical Review Letters}\ }\textbf {\bibinfo {volume} {110}},\ \bibinfo {pages} {133001} (\bibinfo {year} {2013})}\BibitemShut {NoStop}%
\bibitem [{\citenamefont {Kaufman}\ \emph {et~al.}(2012)\citenamefont {Kaufman}, \citenamefont {Lester},\ and\ \citenamefont {Regal}}]{Kaufman2012}%
  \BibitemOpen
  \bibfield  {author} {\bibinfo {author} {\bibfnamefont {A.~M.}\ \bibnamefont {Kaufman}}, \bibinfo {author} {\bibfnamefont {B.~J.}\ \bibnamefont {Lester}},\ and\ \bibinfo {author} {\bibfnamefont {C.~A.}\ \bibnamefont {Regal}},\ }\bibfield  {title} {\bibinfo {title} {Cooling a single atom in an optical tweezer to its quantum ground state},\ }\href {https://doi.org/10.1103/physrevx.2.041014} {\bibfield  {journal} {\bibinfo  {journal} {Physical Review X}\ }\textbf {\bibinfo {volume} {2}},\ \bibinfo {pages} {041014} (\bibinfo {year} {2012})}\BibitemShut {NoStop}%
\bibitem [{\citenamefont {Young}\ \emph {et~al.}(2022)\citenamefont {Young}, \citenamefont {Eckner}, \citenamefont {Schine}, \citenamefont {Childs},\ and\ \citenamefont {Kaufman}}]{Young2022}%
  \BibitemOpen
  \bibfield  {author} {\bibinfo {author} {\bibfnamefont {A.~W.}\ \bibnamefont {Young}}, \bibinfo {author} {\bibfnamefont {W.~J.}\ \bibnamefont {Eckner}}, \bibinfo {author} {\bibfnamefont {N.}~\bibnamefont {Schine}}, \bibinfo {author} {\bibfnamefont {A.~M.}\ \bibnamefont {Childs}},\ and\ \bibinfo {author} {\bibfnamefont {A.~M.}\ \bibnamefont {Kaufman}},\ }\bibfield  {title} {\bibinfo {title} {Tweezer-programmable 2d quantum walks in a hubbard-regime lattice},\ }\href {https://doi.org/10.1126/science.abo0608} {\bibfield  {journal} {\bibinfo  {journal} {Science}\ }\textbf {\bibinfo {volume} {377}},\ \bibinfo {pages} {885} (\bibinfo {year} {2022})}\BibitemShut {NoStop}%
\bibitem [{\citenamefont {Spar}\ \emph {et~al.}(2022)\citenamefont {Spar}, \citenamefont {Guardado-Sanchez}, \citenamefont {Chi}, \citenamefont {Yan},\ and\ \citenamefont {Bakr}}]{Spar2022}%
  \BibitemOpen
  \bibfield  {author} {\bibinfo {author} {\bibfnamefont {B.~M.}\ \bibnamefont {Spar}}, \bibinfo {author} {\bibfnamefont {E.}~\bibnamefont {Guardado-Sanchez}}, \bibinfo {author} {\bibfnamefont {S.}~\bibnamefont {Chi}}, \bibinfo {author} {\bibfnamefont {Z.~Z.}\ \bibnamefont {Yan}},\ and\ \bibinfo {author} {\bibfnamefont {W.~S.}\ \bibnamefont {Bakr}},\ }\bibfield  {title} {\bibinfo {title} {Realization of a fermi-hubbard optical tweezer array},\ }\href {https://doi.org/10.1103/physrevlett.128.223202} {\bibfield  {journal} {\bibinfo  {journal} {Physical Review Letters}\ }\textbf {\bibinfo {volume} {128}},\ \bibinfo {pages} {223202} (\bibinfo {year} {2022})}\BibitemShut {NoStop}%
\bibitem [{\citenamefont {Chew}\ \emph {et~al.}(2022)\citenamefont {Chew}, \citenamefont {Tomita}, \citenamefont {Mahesh}, \citenamefont {Sugawa}, \citenamefont {de~L{\'{e}}s{\'{e}}leuc},\ and\ \citenamefont {Ohmori}}]{Chew2022}%
  \BibitemOpen
  \bibfield  {author} {\bibinfo {author} {\bibfnamefont {Y.}~\bibnamefont {Chew}}, \bibinfo {author} {\bibfnamefont {T.}~\bibnamefont {Tomita}}, \bibinfo {author} {\bibfnamefont {T.~P.}\ \bibnamefont {Mahesh}}, \bibinfo {author} {\bibfnamefont {S.}~\bibnamefont {Sugawa}}, \bibinfo {author} {\bibfnamefont {S.}~\bibnamefont {de~L{\'{e}}s{\'{e}}leuc}},\ and\ \bibinfo {author} {\bibfnamefont {K.}~\bibnamefont {Ohmori}},\ }\bibfield  {title} {\bibinfo {title} {Ultrafast energy exchange between two single rydberg atoms on a nanosecond timescale},\ }\href {https://doi.org/10.1038/s41566-022-01047-2} {\bibfield  {journal} {\bibinfo  {journal} {Nature Photonics}\ }\textbf {\bibinfo {volume} {16}},\ \bibinfo {pages} {724} (\bibinfo {year} {2022})}\BibitemShut {NoStop}%
\bibitem [{\citenamefont {Cooper}\ \emph {et~al.}(2018)\citenamefont {Cooper}, \citenamefont {Covey}, \citenamefont {Madjarov}, \citenamefont {Porsev}, \citenamefont {Safronova},\ and\ \citenamefont {Endres}}]{Cooper2018}%
  \BibitemOpen
  \bibfield  {author} {\bibinfo {author} {\bibfnamefont {A.}~\bibnamefont {Cooper}}, \bibinfo {author} {\bibfnamefont {J.~P.}\ \bibnamefont {Covey}}, \bibinfo {author} {\bibfnamefont {I.~S.}\ \bibnamefont {Madjarov}}, \bibinfo {author} {\bibfnamefont {S.~G.}\ \bibnamefont {Porsev}}, \bibinfo {author} {\bibfnamefont {M.~S.}\ \bibnamefont {Safronova}},\ and\ \bibinfo {author} {\bibfnamefont {M.}~\bibnamefont {Endres}},\ }\bibfield  {title} {\bibinfo {title} {Alkaline-earth atoms in optical tweezers},\ }\href {https://doi.org/10.1103/physrevx.8.041055} {\bibfield  {journal} {\bibinfo  {journal} {Physical Review X}\ }\textbf {\bibinfo {volume} {8}},\ \bibinfo {pages} {041055} (\bibinfo {year} {2018})}\BibitemShut {NoStop}%
\bibitem [{\citenamefont {Norcia}\ \emph {et~al.}(2018)\citenamefont {Norcia}, \citenamefont {Young},\ and\ \citenamefont {Kaufman}}]{Norcia2018}%
  \BibitemOpen
  \bibfield  {author} {\bibinfo {author} {\bibfnamefont {M.}~\bibnamefont {Norcia}}, \bibinfo {author} {\bibfnamefont {A.}~\bibnamefont {Young}},\ and\ \bibinfo {author} {\bibfnamefont {A.}~\bibnamefont {Kaufman}},\ }\bibfield  {title} {\bibinfo {title} {Microscopic control and detection of ultracold strontium in optical-tweezer arrays},\ }\href {https://doi.org/10.1103/physrevx.8.041054} {\bibfield  {journal} {\bibinfo  {journal} {Physical Review X}\ }\textbf {\bibinfo {volume} {8}},\ \bibinfo {pages} {041054} (\bibinfo {year} {2018})}\BibitemShut {NoStop}%
\bibitem [{\citenamefont {Saskin}\ \emph {et~al.}(2019)\citenamefont {Saskin}, \citenamefont {Wilson}, \citenamefont {Grinkemeyer},\ and\ \citenamefont {Thompson}}]{Saskin2019}%
  \BibitemOpen
  \bibfield  {author} {\bibinfo {author} {\bibfnamefont {S.}~\bibnamefont {Saskin}}, \bibinfo {author} {\bibfnamefont {J.}~\bibnamefont {Wilson}}, \bibinfo {author} {\bibfnamefont {B.}~\bibnamefont {Grinkemeyer}},\ and\ \bibinfo {author} {\bibfnamefont {J.}~\bibnamefont {Thompson}},\ }\bibfield  {title} {\bibinfo {title} {Narrow-line cooling and imaging of ytterbium atoms in an optical tweezer array},\ }\href {https://doi.org/10.1103/physrevlett.122.143002} {\bibfield  {journal} {\bibinfo  {journal} {Physical Review Letters}\ }\textbf {\bibinfo {volume} {122}},\ \bibinfo {pages} {143002} (\bibinfo {year} {2019})}\BibitemShut {NoStop}%
\bibitem [{\citenamefont {Meinert}\ \emph {et~al.}(2021)\citenamefont {Meinert}, \citenamefont {Pfau},\ and\ \citenamefont {H\"{o}lzl}}]{PatentFineStructureQubit}%
  \BibitemOpen
  \bibfield  {author} {\bibinfo {author} {\bibfnamefont {F.}~\bibnamefont {Meinert}}, \bibinfo {author} {\bibfnamefont {T.}~\bibnamefont {Pfau}},\ and\ \bibinfo {author} {\bibfnamefont {C.}~\bibnamefont {H\"{o}lzl}},\ }\href@noop {} {\bibinfo {title} {{Quantum computing device, use, and method, European Patent Application EP20214187.5}}} (\bibinfo {year} {2021})\BibitemShut {NoStop}%
\bibitem [{\citenamefont {Pagano}\ \emph {et~al.}(2022)\citenamefont {Pagano}, \citenamefont {Weber}, \citenamefont {Jaschke}, \citenamefont {Pfau}, \citenamefont {Meinert}, \citenamefont {Montangero},\ and\ \citenamefont {B\"{u}chler}}]{Pagano2022}%
  \BibitemOpen
  \bibfield  {author} {\bibinfo {author} {\bibfnamefont {A.}~\bibnamefont {Pagano}}, \bibinfo {author} {\bibfnamefont {S.}~\bibnamefont {Weber}}, \bibinfo {author} {\bibfnamefont {D.}~\bibnamefont {Jaschke}}, \bibinfo {author} {\bibfnamefont {T.}~\bibnamefont {Pfau}}, \bibinfo {author} {\bibfnamefont {F.}~\bibnamefont {Meinert}}, \bibinfo {author} {\bibfnamefont {S.}~\bibnamefont {Montangero}},\ and\ \bibinfo {author} {\bibfnamefont {H.~P.}\ \bibnamefont {B\"{u}chler}},\ }\bibfield  {title} {\bibinfo {title} {Error budgeting for a controlled-phase gate with strontium-88 rydberg atoms},\ }\href {https://doi.org/10.1103/physrevresearch.4.033019} {\bibfield  {journal} {\bibinfo  {journal} {Physical Review Research}\ }\textbf {\bibinfo {volume} {4}},\ \bibinfo {pages} {033019} (\bibinfo {year} {2022})}\BibitemShut {NoStop}%
\bibitem [{\citenamefont {Anderegg}\ \emph {et~al.}(2019)\citenamefont {Anderegg}, \citenamefont {Cheuk}, \citenamefont {Bao}, \citenamefont {Burchesky}, \citenamefont {Ketterle}, \citenamefont {Ni},\ and\ \citenamefont {Doyle}}]{Anderegg2019}%
  \BibitemOpen
  \bibfield  {author} {\bibinfo {author} {\bibfnamefont {L.}~\bibnamefont {Anderegg}}, \bibinfo {author} {\bibfnamefont {L.~W.}\ \bibnamefont {Cheuk}}, \bibinfo {author} {\bibfnamefont {Y.}~\bibnamefont {Bao}}, \bibinfo {author} {\bibfnamefont {S.}~\bibnamefont {Burchesky}}, \bibinfo {author} {\bibfnamefont {W.}~\bibnamefont {Ketterle}}, \bibinfo {author} {\bibfnamefont {K.-K.}\ \bibnamefont {Ni}},\ and\ \bibinfo {author} {\bibfnamefont {J.~M.}\ \bibnamefont {Doyle}},\ }\bibfield  {title} {\bibinfo {title} {An optical tweezer array of ultracold molecules},\ }\href {https://doi.org/10.1126/science.aax1265} {\bibfield  {journal} {\bibinfo  {journal} {Science}\ }\textbf {\bibinfo {volume} {365}},\ \bibinfo {pages} {1156} (\bibinfo {year} {2019})}\BibitemShut {NoStop}%
\bibitem [{\citenamefont {Caldwell}\ and\ \citenamefont {Tarbutt}(2020)}]{Caldwell2020}%
  \BibitemOpen
  \bibfield  {author} {\bibinfo {author} {\bibfnamefont {L.}~\bibnamefont {Caldwell}}\ and\ \bibinfo {author} {\bibfnamefont {M.~R.}\ \bibnamefont {Tarbutt}},\ }\bibfield  {title} {\bibinfo {title} {Sideband cooling of molecules in optical traps},\ }\href {https://doi.org/10.1103/physrevresearch.2.013251} {\bibfield  {journal} {\bibinfo  {journal} {Physical Review Research}\ }\textbf {\bibinfo {volume} {2}},\ \bibinfo {pages} {013251} (\bibinfo {year} {2020})}\BibitemShut {NoStop}%
\bibitem [{\citenamefont {Schneider}\ \emph {et~al.}(2010)\citenamefont {Schneider}, \citenamefont {Enderlein}, \citenamefont {Huber},\ and\ \citenamefont {Schaetz}}]{Schneider2010}%
  \BibitemOpen
  \bibfield  {author} {\bibinfo {author} {\bibfnamefont {C.}~\bibnamefont {Schneider}}, \bibinfo {author} {\bibfnamefont {M.}~\bibnamefont {Enderlein}}, \bibinfo {author} {\bibfnamefont {T.}~\bibnamefont {Huber}},\ and\ \bibinfo {author} {\bibfnamefont {T.}~\bibnamefont {Schaetz}},\ }\bibfield  {title} {\bibinfo {title} {Optical trapping of an ion},\ }\href {https://doi.org/10.1038/nphoton.2010.236} {\bibfield  {journal} {\bibinfo  {journal} {Nature Photonics}\ }\textbf {\bibinfo {volume} {4}},\ \bibinfo {pages} {772} (\bibinfo {year} {2010})}\BibitemShut {NoStop}%
\bibitem [{\citenamefont {Berto}\ \emph {et~al.}(2021)\citenamefont {Berto}, \citenamefont {Perego}, \citenamefont {Duca},\ and\ \citenamefont {Sias}}]{Berto2021}%
  \BibitemOpen
  \bibfield  {author} {\bibinfo {author} {\bibfnamefont {F.}~\bibnamefont {Berto}}, \bibinfo {author} {\bibfnamefont {E.}~\bibnamefont {Perego}}, \bibinfo {author} {\bibfnamefont {L.}~\bibnamefont {Duca}},\ and\ \bibinfo {author} {\bibfnamefont {C.}~\bibnamefont {Sias}},\ }\bibfield  {title} {\bibinfo {title} {Prospects for single-photon sideband cooling of optically trapped neutral atoms},\ }\href {https://doi.org/10.1103/physrevresearch.3.043106} {\bibfield  {journal} {\bibinfo  {journal} {Physical Review Research}\ }\textbf {\bibinfo {volume} {3}},\ \bibinfo {pages} {043106} (\bibinfo {year} {2021})}\BibitemShut {NoStop}%
\bibitem [{\citenamefont {Endres}\ \emph {et~al.}(2016)\citenamefont {Endres}, \citenamefont {Bernien}, \citenamefont {Keesling}, \citenamefont {Levine}, \citenamefont {Anschuetz}, \citenamefont {Krajenbrink}, \citenamefont {Senko}, \citenamefont {Vuletic}, \citenamefont {Greiner},\ and\ \citenamefont {Lukin}}]{Endres2016}%
  \BibitemOpen
  \bibfield  {author} {\bibinfo {author} {\bibfnamefont {M.}~\bibnamefont {Endres}}, \bibinfo {author} {\bibfnamefont {H.}~\bibnamefont {Bernien}}, \bibinfo {author} {\bibfnamefont {A.}~\bibnamefont {Keesling}}, \bibinfo {author} {\bibfnamefont {H.}~\bibnamefont {Levine}}, \bibinfo {author} {\bibfnamefont {E.~R.}\ \bibnamefont {Anschuetz}}, \bibinfo {author} {\bibfnamefont {A.}~\bibnamefont {Krajenbrink}}, \bibinfo {author} {\bibfnamefont {C.}~\bibnamefont {Senko}}, \bibinfo {author} {\bibfnamefont {V.}~\bibnamefont {Vuletic}}, \bibinfo {author} {\bibfnamefont {M.}~\bibnamefont {Greiner}},\ and\ \bibinfo {author} {\bibfnamefont {M.~D.}\ \bibnamefont {Lukin}},\ }\bibfield  {title} {\bibinfo {title} {Atom-by-atom assembly of defect-free one-dimensional cold atom arrays},\ }\href {https://doi.org/10.1126/science.aah3752} {\bibfield  {journal} {\bibinfo  {journal} {Science}\ }\textbf {\bibinfo {volume} {354}},\ \bibinfo {pages} {1024} (\bibinfo {year} {2016})}\BibitemShut {NoStop}%
\bibitem [{\citenamefont {Barredo}\ \emph {et~al.}(2016)\citenamefont {Barredo}, \citenamefont {de~L{\'{e}}s{\'{e}}leuc}, \citenamefont {Lienhard}, \citenamefont {Lahaye},\ and\ \citenamefont {Browaeys}}]{Barredo2016}%
  \BibitemOpen
  \bibfield  {author} {\bibinfo {author} {\bibfnamefont {D.}~\bibnamefont {Barredo}}, \bibinfo {author} {\bibfnamefont {S.}~\bibnamefont {de~L{\'{e}}s{\'{e}}leuc}}, \bibinfo {author} {\bibfnamefont {V.}~\bibnamefont {Lienhard}}, \bibinfo {author} {\bibfnamefont {T.}~\bibnamefont {Lahaye}},\ and\ \bibinfo {author} {\bibfnamefont {A.}~\bibnamefont {Browaeys}},\ }\bibfield  {title} {\bibinfo {title} {An atom-by-atom assembler of defect-free arbitrary two-dimensional atomic arrays},\ }\href {https://doi.org/10.1126/science.aah3778} {\bibfield  {journal} {\bibinfo  {journal} {Science}\ }\textbf {\bibinfo {volume} {354}},\ \bibinfo {pages} {1021} (\bibinfo {year} {2016})}\BibitemShut {NoStop}%
\bibitem [{\citenamefont {Santos}\ \emph {et~al.}(2000)\citenamefont {Santos}, \citenamefont {Idziaszek}, \citenamefont {Cirac},\ and\ \citenamefont {Lewenstein}}]{Santos2000}%
  \BibitemOpen
  \bibfield  {author} {\bibinfo {author} {\bibfnamefont {L.}~\bibnamefont {Santos}}, \bibinfo {author} {\bibfnamefont {Z.}~\bibnamefont {Idziaszek}}, \bibinfo {author} {\bibfnamefont {J.~I.}\ \bibnamefont {Cirac}},\ and\ \bibinfo {author} {\bibfnamefont {M.}~\bibnamefont {Lewenstein}},\ }\bibfield  {title} {\bibinfo {title} {Laser-induced condensation of trapped bosonic gases},\ }\href {https://doi.org/10.1088/0953-4075/33/19/322} {\bibfield  {journal} {\bibinfo  {journal} {Journal of Physics B: Atomic, Molecular and Optical Physics}\ }\textbf {\bibinfo {volume} {33}},\ \bibinfo {pages} {4131} (\bibinfo {year} {2000})}\BibitemShut {NoStop}%
\bibitem [{\citenamefont {Urvoy}\ \emph {et~al.}(2019)\citenamefont {Urvoy}, \citenamefont {Vendeiro}, \citenamefont {Ramette}, \citenamefont {Adiyatullin},\ and\ \citenamefont {Vuleti{\'{c}}}}]{Urvoy2019}%
  \BibitemOpen
  \bibfield  {author} {\bibinfo {author} {\bibfnamefont {A.}~\bibnamefont {Urvoy}}, \bibinfo {author} {\bibfnamefont {Z.}~\bibnamefont {Vendeiro}}, \bibinfo {author} {\bibfnamefont {J.}~\bibnamefont {Ramette}}, \bibinfo {author} {\bibfnamefont {A.}~\bibnamefont {Adiyatullin}},\ and\ \bibinfo {author} {\bibfnamefont {V.}~\bibnamefont {Vuleti{\'{c}}}},\ }\bibfield  {title} {\bibinfo {title} {Direct laser cooling to bose-einstein condensation in a dipole trap},\ }\href {https://doi.org/10.1103/physrevlett.122.203202} {\bibfield  {journal} {\bibinfo  {journal} {Physical Review Letters}\ }\textbf {\bibinfo {volume} {122}},\ \bibinfo {pages} {203202} (\bibinfo {year} {2019})}\BibitemShut {NoStop}%
\bibitem [{\citenamefont {Covey}\ \emph {et~al.}(2019)\citenamefont {Covey}, \citenamefont {Madjarov}, \citenamefont {Cooper},\ and\ \citenamefont {Endres}}]{Covey2019}%
  \BibitemOpen
  \bibfield  {author} {\bibinfo {author} {\bibfnamefont {J.~P.}\ \bibnamefont {Covey}}, \bibinfo {author} {\bibfnamefont {I.~S.}\ \bibnamefont {Madjarov}}, \bibinfo {author} {\bibfnamefont {A.}~\bibnamefont {Cooper}},\ and\ \bibinfo {author} {\bibfnamefont {M.}~\bibnamefont {Endres}},\ }\bibfield  {title} {\bibinfo {title} {2000-times repeated imaging of strontium atoms in clock-magic tweezer arrays},\ }\href {https://doi.org/10.1103/physrevlett.122.173201} {\bibfield  {journal} {\bibinfo  {journal} {Physical Review Letters}\ }\textbf {\bibinfo {volume} {122}},\ \bibinfo {pages} {173201} (\bibinfo {year} {2019})}\BibitemShut {NoStop}%
\bibitem [{\citenamefont {Urech}\ \emph {et~al.}(2022)\citenamefont {Urech}, \citenamefont {Knottnerus}, \citenamefont {Spreeuw},\ and\ \citenamefont {Schreck}}]{Urech2022}%
  \BibitemOpen
  \bibfield  {author} {\bibinfo {author} {\bibfnamefont {A.}~\bibnamefont {Urech}}, \bibinfo {author} {\bibfnamefont {I.~H.~A.}\ \bibnamefont {Knottnerus}}, \bibinfo {author} {\bibfnamefont {R.~J.~C.}\ \bibnamefont {Spreeuw}},\ and\ \bibinfo {author} {\bibfnamefont {F.}~\bibnamefont {Schreck}},\ }\bibfield  {title} {\bibinfo {title} {Narrow-line imaging of single strontium atoms in shallow optical tweezers},\ }\href {https://doi.org/10.1103/physrevresearch.4.023245} {\bibfield  {journal} {\bibinfo  {journal} {Physical Review Research}\ }\textbf {\bibinfo {volume} {4}},\ \bibinfo {pages} {023245} (\bibinfo {year} {2022})}\BibitemShut {NoStop}%
\bibitem [{\citenamefont {Ta{\"i}eb}\ \emph {et~al.}(1994)\citenamefont {Ta{\"i}eb}, \citenamefont {Dum}, \citenamefont {Cirac}, \citenamefont {Marte},\ and\ \citenamefont {Zoller}}]{Taieb1994}%
  \BibitemOpen
  \bibfield  {author} {\bibinfo {author} {\bibfnamefont {R.}~\bibnamefont {Ta{\"i}eb}}, \bibinfo {author} {\bibfnamefont {R.}~\bibnamefont {Dum}}, \bibinfo {author} {\bibfnamefont {J.~I.}\ \bibnamefont {Cirac}}, \bibinfo {author} {\bibfnamefont {P.}~\bibnamefont {Marte}},\ and\ \bibinfo {author} {\bibfnamefont {P.}~\bibnamefont {Zoller}},\ }\bibfield  {title} {\bibinfo {title} {Cooling and localization of atoms in laser-induced potential wells},\ }\href {https://doi.org/10.1103/physreva.49.4876} {\bibfield  {journal} {\bibinfo  {journal} {Physical Review A}\ }\textbf {\bibinfo {volume} {49}},\ \bibinfo {pages} {4876} (\bibinfo {year} {1994})}\BibitemShut {NoStop}%
\bibitem [{\citenamefont {Tuchendler}\ \emph {et~al.}(2008)\citenamefont {Tuchendler}, \citenamefont {Lance}, \citenamefont {Browaeys}, \citenamefont {Sortais},\ and\ \citenamefont {Grangier}}]{Tuchendler2008}%
  \BibitemOpen
  \bibfield  {author} {\bibinfo {author} {\bibfnamefont {C.}~\bibnamefont {Tuchendler}}, \bibinfo {author} {\bibfnamefont {A.~M.}\ \bibnamefont {Lance}}, \bibinfo {author} {\bibfnamefont {A.}~\bibnamefont {Browaeys}}, \bibinfo {author} {\bibfnamefont {Y.~R.~P.}\ \bibnamefont {Sortais}},\ and\ \bibinfo {author} {\bibfnamefont {P.}~\bibnamefont {Grangier}},\ }\bibfield  {title} {\bibinfo {title} {Energy distribution and cooling of a single atom in an optical tweezer},\ }\href {https://doi.org/10.1103/physreva.78.033425} {\bibfield  {journal} {\bibinfo  {journal} {Physical Review A}\ }\textbf {\bibinfo {volume} {78}},\ \bibinfo {pages} {033425} (\bibinfo {year} {2008})}\BibitemShut {NoStop}%
\bibitem [{\citenamefont {Schlosser}\ \emph {et~al.}(2002)\citenamefont {Schlosser}, \citenamefont {Reymond},\ and\ \citenamefont {Grangier}}]{Schlosser2002}%
  \BibitemOpen
  \bibfield  {author} {\bibinfo {author} {\bibfnamefont {N.}~\bibnamefont {Schlosser}}, \bibinfo {author} {\bibfnamefont {G.}~\bibnamefont {Reymond}},\ and\ \bibinfo {author} {\bibfnamefont {P.}~\bibnamefont {Grangier}},\ }\bibfield  {title} {\bibinfo {title} {Collisional blockade in microscopic optical dipole traps},\ }\href {https://doi.org/10.1103/physrevlett.89.023005} {\bibfield  {journal} {\bibinfo  {journal} {Physical Review Letters}\ }\textbf {\bibinfo {volume} {89}},\ \bibinfo {pages} {023005} (\bibinfo {year} {2002})}\BibitemShut {NoStop}%
\bibitem [{\citenamefont {Gr\"{u}nzweig}\ \emph {et~al.}(2010)\citenamefont {Gr\"{u}nzweig}, \citenamefont {Hilliard}, \citenamefont {McGovern},\ and\ \citenamefont {Andersen}}]{Gruenzweig2010}%
  \BibitemOpen
  \bibfield  {author} {\bibinfo {author} {\bibfnamefont {T.}~\bibnamefont {Gr\"{u}nzweig}}, \bibinfo {author} {\bibfnamefont {A.}~\bibnamefont {Hilliard}}, \bibinfo {author} {\bibfnamefont {M.}~\bibnamefont {McGovern}},\ and\ \bibinfo {author} {\bibfnamefont {M.~F.}\ \bibnamefont {Andersen}},\ }\bibfield  {title} {\bibinfo {title} {Near-deterministic preparation of a single atom in an optical microtrap},\ }\href {https://doi.org/10.1038/nphys1778} {\bibfield  {journal} {\bibinfo  {journal} {Nature Physics}\ }\textbf {\bibinfo {volume} {6}},\ \bibinfo {pages} {951} (\bibinfo {year} {2010})}\BibitemShut {NoStop}%
\bibitem [{\citenamefont {Zelevinsky}\ \emph {et~al.}(2006)\citenamefont {Zelevinsky}, \citenamefont {Boyd}, \citenamefont {Ludlow}, \citenamefont {Ido}, \citenamefont {Ye}, \citenamefont {Ciury{\l}o}, \citenamefont {Naidon},\ and\ \citenamefont {Julienne}}]{Zelevinsky2006}%
  \BibitemOpen
  \bibfield  {author} {\bibinfo {author} {\bibfnamefont {T.}~\bibnamefont {Zelevinsky}}, \bibinfo {author} {\bibfnamefont {M.~M.}\ \bibnamefont {Boyd}}, \bibinfo {author} {\bibfnamefont {A.~D.}\ \bibnamefont {Ludlow}}, \bibinfo {author} {\bibfnamefont {T.}~\bibnamefont {Ido}}, \bibinfo {author} {\bibfnamefont {J.}~\bibnamefont {Ye}}, \bibinfo {author} {\bibfnamefont {R.}~\bibnamefont {Ciury{\l}o}}, \bibinfo {author} {\bibfnamefont {P.}~\bibnamefont {Naidon}},\ and\ \bibinfo {author} {\bibfnamefont {P.~S.}\ \bibnamefont {Julienne}},\ }\bibfield  {title} {\bibinfo {title} {Narrow line photoassociation in an optical lattice},\ }\href {https://doi.org/10.1103/physrevlett.96.203201} {\bibfield  {journal} {\bibinfo  {journal} {Physical Review Letters}\ }\textbf {\bibinfo {volume} {96}},\ \bibinfo {pages} {203201} (\bibinfo {year} {2006})}\BibitemShut {NoStop}%
\bibitem [{\citenamefont {Barnes}\ \emph {et~al.}(2022)\citenamefont {Barnes}, \citenamefont {Battaglino}, \citenamefont {Bloom}, \citenamefont {Cassella}, \citenamefont {Coxe}, \citenamefont {Crisosto}, \citenamefont {King}, \citenamefont {Kondov}, \citenamefont {Kotru}, \citenamefont {Larsen}, \citenamefont {Lauigan}, \citenamefont {Lester}, \citenamefont {McDonald}, \citenamefont {Megidish}, \citenamefont {Narayanaswami}, \citenamefont {Nishiguchi}, \citenamefont {Notermans}, \citenamefont {Peng}, \citenamefont {Ryou}, \citenamefont {Wu},\ and\ \citenamefont {Yarwood}}]{Barnes2022}%
  \BibitemOpen
  \bibfield  {author} {\bibinfo {author} {\bibfnamefont {K.}~\bibnamefont {Barnes}}, \bibinfo {author} {\bibfnamefont {P.}~\bibnamefont {Battaglino}}, \bibinfo {author} {\bibfnamefont {B.~J.}\ \bibnamefont {Bloom}}, \bibinfo {author} {\bibfnamefont {K.}~\bibnamefont {Cassella}}, \bibinfo {author} {\bibfnamefont {R.}~\bibnamefont {Coxe}}, \bibinfo {author} {\bibfnamefont {N.}~\bibnamefont {Crisosto}}, \bibinfo {author} {\bibfnamefont {J.~P.}\ \bibnamefont {King}}, \bibinfo {author} {\bibfnamefont {S.~S.}\ \bibnamefont {Kondov}}, \bibinfo {author} {\bibfnamefont {K.}~\bibnamefont {Kotru}}, \bibinfo {author} {\bibfnamefont {S.~C.}\ \bibnamefont {Larsen}}, \bibinfo {author} {\bibfnamefont {J.}~\bibnamefont {Lauigan}}, \bibinfo {author} {\bibfnamefont {B.~J.}\ \bibnamefont {Lester}}, \bibinfo {author} {\bibfnamefont {M.}~\bibnamefont {McDonald}}, \bibinfo {author} {\bibfnamefont {E.}~\bibnamefont {Megidish}}, \bibinfo {author} {\bibfnamefont {S.}~\bibnamefont {Narayanaswami}}, \bibinfo {author} {\bibfnamefont {C.}~\bibnamefont {Nishiguchi}}, \bibinfo {author} {\bibfnamefont {R.}~\bibnamefont {Notermans}}, \bibinfo {author} {\bibfnamefont {L.~S.}\ \bibnamefont {Peng}}, \bibinfo {author} {\bibfnamefont {A.}~\bibnamefont {Ryou}}, \bibinfo {author} {\bibfnamefont {T.-Y.}\ \bibnamefont {Wu}},\ and\ \bibinfo {author} {\bibfnamefont {M.}~\bibnamefont {Yarwood}},\ }\bibfield  {title} {\bibinfo {title} {Assembly and coherent control of a register of nuclear spin qubits},\ }\href {https://doi.org/10.1038/s41467-022-29977-z} {\bibfield  {journal} {\bibinfo  {journal} {Nature Communications}\ }\textbf {\bibinfo {volume} {13}},\ \bibinfo {pages} {2779} (\bibinfo {year} {2022})}\BibitemShut {NoStop}%
\bibitem [{\citenamefont {Kien}\ \emph {et~al.}(2013)\citenamefont {Kien}, \citenamefont {Schneeweiss},\ and\ \citenamefont {Rauschenbeutel}}]{Kien2013}%
  \BibitemOpen
  \bibfield  {author} {\bibinfo {author} {\bibfnamefont {F.~L.}\ \bibnamefont {Kien}}, \bibinfo {author} {\bibfnamefont {P.}~\bibnamefont {Schneeweiss}},\ and\ \bibinfo {author} {\bibfnamefont {A.}~\bibnamefont {Rauschenbeutel}},\ }\bibfield  {title} {\bibinfo {title} {Dynamical polarizability of atoms in arbitrary light fields: general theory and application to cesium},\ }\href {https://doi.org/10.1140/epjd/e2013-30729-x} {\bibfield  {journal} {\bibinfo  {journal} {The European Physical Journal D}\ }\textbf {\bibinfo {volume} {67}},\ \bibinfo {pages} {92} (\bibinfo {year} {2013})}\BibitemShut {NoStop}%
\bibitem [{\citenamefont {Kozlov}\ and\ \citenamefont {Porsev}(1999)}]{Kozlov1999}%
  \BibitemOpen
  \bibfield  {author} {\bibinfo {author} {\bibfnamefont {M.}~\bibnamefont {Kozlov}}\ and\ \bibinfo {author} {\bibfnamefont {S.}~\bibnamefont {Porsev}},\ }\bibfield  {title} {\bibinfo {title} {Polarizabilities and hyperfine structure constants of the low-lying levels of barium},\ }\href {https://doi.org/10.1007/s100530050229} {\bibfield  {journal} {\bibinfo  {journal} {The European Physical Journal D - Atomic, Molecular and Optical Physics}\ }\textbf {\bibinfo {volume} {5}},\ \bibinfo {pages} {59} (\bibinfo {year} {1999})}\BibitemShut {NoStop}%
\bibitem [{\citenamefont {Dalgarno}\ and\ \citenamefont {Lewis}(1955)}]{Dalgarno1955}%
  \BibitemOpen
  \bibfield  {author} {\bibinfo {author} {\bibfnamefont {A.}~\bibnamefont {Dalgarno}}\ and\ \bibinfo {author} {\bibfnamefont {J.~T.}\ \bibnamefont {Lewis}},\ }\bibfield  {title} {\bibinfo {title} {The exact calculation of long-range forces between atoms by perturbation theory},\ }\href {https://doi.org/10.1098/rspa.1955.0246} {\bibfield  {journal} {\bibinfo  {journal} {Proceedings of the Royal Society of London. Series A. Mathematical and Physical Sciences}\ }\textbf {\bibinfo {volume} {233}},\ \bibinfo {pages} {70} (\bibinfo {year} {1955})}\BibitemShut {NoStop}%
\bibitem [{\citenamefont {Cheung}\ \emph {et~al.}(2021)\citenamefont {Cheung}, \citenamefont {Safronova},\ and\ \citenamefont {Porsev}}]{Cheung2021}%
  \BibitemOpen
  \bibfield  {author} {\bibinfo {author} {\bibfnamefont {C.}~\bibnamefont {Cheung}}, \bibinfo {author} {\bibfnamefont {M.}~\bibnamefont {Safronova}},\ and\ \bibinfo {author} {\bibfnamefont {S.}~\bibnamefont {Porsev}},\ }\bibfield  {title} {\bibinfo {title} {Scalable codes for precision calculations of properties of complex atomic systems},\ }\href {https://doi.org/10.3390/sym13040621} {\bibfield  {journal} {\bibinfo  {journal} {Symmetry}\ }\textbf {\bibinfo {volume} {13}},\ \bibinfo {pages} {621} (\bibinfo {year} {2021})}\BibitemShut {NoStop}%
\bibitem [{\citenamefont {Safronova}\ \emph {et~al.}(2009)\citenamefont {Safronova}, \citenamefont {Kozlov}, \citenamefont {Johnson},\ and\ \citenamefont {Jiang}}]{Safronova2009}%
  \BibitemOpen
  \bibfield  {author} {\bibinfo {author} {\bibfnamefont {M.~S.}\ \bibnamefont {Safronova}}, \bibinfo {author} {\bibfnamefont {M.~G.}\ \bibnamefont {Kozlov}}, \bibinfo {author} {\bibfnamefont {W.~R.}\ \bibnamefont {Johnson}},\ and\ \bibinfo {author} {\bibfnamefont {D.}~\bibnamefont {Jiang}},\ }\bibfield  {title} {\bibinfo {title} {Development of a configuration-interaction plus all-order method for atomic calculations},\ }\href {https://doi.org/10.1103/physreva.80.012516} {\bibfield  {journal} {\bibinfo  {journal} {Physical Review A}\ }\textbf {\bibinfo {volume} {80}},\ \bibinfo {pages} {012516} (\bibinfo {year} {2009})}\BibitemShut {NoStop}%
\bibitem [{\citenamefont {Kale}(2020)}]{Kale2020}%
  \BibitemOpen
  \bibfield  {author} {\bibinfo {author} {\bibfnamefont {A.~M.}\ \bibnamefont {Kale}},\ }\bibfield  {title} {\bibinfo {title} {Towards high fidelity quantum computation and simulation with rydberg atoms},\ }\href {https://doi.org/10.7907/8MEE-MD98} {(\bibinfo {year} {2020})}\BibitemShut {NoStop}%
\bibitem [{\citenamefont {Johansson}\ \emph {et~al.}(2013)\citenamefont {Johansson}, \citenamefont {Nation},\ and\ \citenamefont {Nori}}]{Johansson2013}%
  \BibitemOpen
  \bibfield  {author} {\bibinfo {author} {\bibfnamefont {J.}~\bibnamefont {Johansson}}, \bibinfo {author} {\bibfnamefont {P.}~\bibnamefont {Nation}},\ and\ \bibinfo {author} {\bibfnamefont {F.}~\bibnamefont {Nori}},\ }\bibfield  {title} {\bibinfo {title} {{QuTiP} 2: A python framework for the dynamics of open quantum systems},\ }\href {https://doi.org/10.1016/j.cpc.2012.11.019} {\bibfield  {journal} {\bibinfo  {journal} {Computer Physics Communications}\ }\textbf {\bibinfo {volume} {184}},\ \bibinfo {pages} {1234} (\bibinfo {year} {2013})}\BibitemShut {NoStop}%
\end{thebibliography}
%

\end{document}